\documentclass[10pt,a4paper]{article}
\usepackage[latin1]{inputenc}
\usepackage{amsmath}
\usepackage{amsfonts}
\usepackage{amssymb}
\usepackage{graphics}
\usepackage{subfig}
\usepackage{psfrag}
\usepackage{epsfig}
\usepackage{pstricks}
\usepackage{float}
\usepackage{empheq}

\newcommand{\m}[1]{\mathcal{#1}}
\newcommand{\ph}[1]{\phantom{#1}}
\newcommand{\w}{\wedge}

\newcommand{\nn}{\nonumber}
\newcommand{\mb}[1]{\mathbb{#1}}

\author{\\\\
Hans F. Westman$^1$\footnote{\texttt{westman@iff.csic.es}}\; and Tom G. Zlosnik$^2$\footnote{\texttt{tom.zlosnik@gmail.com}}\\
{\small \it $(1)$ Instituto de F\'isica Fundamental, CSIC, Serrano 113-B, 28006 Madrid, Spain}\\
{\small \it $(2)$ Imperial College Theoretical Physics, Huxley Building, London, SW7 2AZ}
}
\date{\today}
\title{Gravity, Cartan geometry, and idealized waywisers}
\begin{document}
\maketitle
\begin{abstract}
The primary aim of this paper is to provide a simple and concrete interpretation of Cartan geometry in terms of the mathematics of idealized waywisers. Waywisers, also called hodometers, are instruments traditionally used to measure distances. The mathematical representation of an idealized waywiser consists of a choice of symmetric space called a {\em model space} and represents the `wheel' of the idealized waywiser. The geometry of a manifold is then completely characterized by a pair of variables $\{V^A(x),A^{AB}(x)\}$, each of which admit simple interpretations: $V^A$ is the point of contact between the waywiser's idealized wheel and the manifold whose geometry one wishes to characterize, and $A^{AB}=A_\mu^{\ph\mu AB}dx^\mu$ is a connection one-form dictating how much the idealized wheel of the waywiser has rotated when rolled along the manifold. The familiar objects from differential geometry (e.g. metric $g_{\mu\nu}$, affine connection $\Gamma^\rho_{\mu\nu}$, co-tetrad $e^I$, torsion $T^I$, spin-connection $\omega^{IJ}$, Riemannian curvature $R^{IJ}$) can be seen as compound objects made out of the waywiser variables $\{V^A,A^{AB}\}$. We then generalize this waywiser approach to relativistic spacetimes and exhibit action principles for General Relativity in terms of the waywiser variables for two choices of model {\em spacetimes}: De Sitter and anti-De Sitter spacetimes.
\end{abstract}

\section{Introduction}\label{intro}
Riemannian geometry forms the mathematical basis of Einstein's General Relativity. The metric representation of Riemannian geometry consists of the pair of variables $\{g_{\mu\nu},\Gamma^\rho_{\mu\nu}\}$. While the symmetric metric tensor $g_{\mu\nu}$ encodes all information of distances between points on a manifold, the affine connection $\Gamma^\rho_{\mu\nu}$ encodes the information of parallel transport of tangent vectors $u^\mu$ as well as defining a covariant derivative $\nabla_\mu$ acting on tensors. Within Riemannian geometry, not all pairs $\{g_{\mu\nu},\Gamma^\rho_{\mu\nu}\}$ are allowed, and two conditions are imposed:
\begin{itemize}
\item {\bf Metric compatibility:} $\nabla_\rho g_{\mu\nu}=\partial_\rho g_{\mu\nu}-\Gamma^\sigma_{\rho\mu}g_{\sigma\nu}-\Gamma^\sigma_{\rho\nu}g_{\mu\sigma}=0$
\item {\bf Zero torsion:} $T^\rho_{\mu\nu}\equiv\Gamma^\rho_{\mu\nu}-\Gamma^\rho_{\nu\mu}=0$.
\end{itemize}
The affine connection can then be uniquely determined from the metric 
\begin{eqnarray}
\Gamma^\rho_{\mu\nu}=\frac{1}{2}g^{\rho\sigma}(\partial_\mu g_{\sigma\nu}+\partial_\nu g_{\mu\sigma}-\partial_\sigma g_{\mu\nu})
\end{eqnarray}
and it becomes natural to view the metric as the primary variable and the affine connection as a secondary derived quantity. Metric-compatibility admits a crisp geometric interpretation: an affine straight line $X^\mu(\lambda)$ (i.e. an affine geodesic) between two points $X^\mu(\lambda_1)=x_1^\mu$ and $X^\mu(\lambda_2)=x_2^\mu$ is also the metrically shortest path (or longest in the case of timelike paths) between two points (and {\em vice versa}),
\begin{eqnarray}
\frac{\delta}{\delta X^\rho}\int\sqrt{g_{\mu\nu}(X(\lambda))\dot{X}^\mu\dot{X}^\nu}d\lambda=0\qquad\Leftrightarrow \qquad\frac{dX^\mu}{d\lambda}+\Gamma^\rho_{\mu\nu}\dot{X}^\mu\dot{X}^\nu\propto \dot{X}^\mu
\end{eqnarray}
here $\dot{X}^\mu\equiv \frac{dX^\mu}{d\lambda}$ and $\frac{\delta}{\delta X}$ denotes a variational derivative with $\delta X(\lambda_1)=\delta X(\lambda_2)=0$. However, this condition does not fix the non-symmetric part of the connection, i.e. the torsion. This is related to the fact that affine geodesics, and consequently also celestial motion, are unaffected by the presence of torsion. 

As is well-known, the existence of fermionic matter in nature has immediate implications for the mathematical representation of the gravitational field. When it comes to coupling a spinor field to the gravitational field it is known that the metric representation (as defined above) is unsuitable. The fundamental reason for this is that a spinor constitutes a finite-dimensional spin-half representation of the Lorentz group while the affine connection is $GL(4)$-valued, a group which admits no finite-dimensional spinorial representation \cite{cartans}. Instead, whenever fermionic matter is present the metric representation is shunned and the gravitational field is instead mathematically represented in terms of a pair of $\mathfrak{so}(1,3)$-valued one-forms: the co-tetrad $e^I=e^I_\mu dx^\mu$ and the spin connection $\omega^I_{\ph IJ}=\omega_{\mu\ph IJ}^{\ph \mu I}dx^\mu$ (see for example \cite{Trautman:2006fp}). Given the notion of the spin connection one-form, one may define a linear covariant exterior derivative of a spinor $\psi$: 
\begin{eqnarray}
D\psi=d\psi-\frac{i}{2}\omega_{IJ}S^{IJ}\psi.
\end{eqnarray}
where $S^{IJ}=-\frac{i}{4}[\gamma^I,\gamma^J]$. On the other hand, a linear covariant derivative cannot be defined for a spinor within the metric representation. Absent a satisfactory solution of the mathematical problem the existence of fermions pose, we must therefore discard the metric representation as a viable mathematical representation of the gravitational field. 

The Einstein-Palatini-Dirac action for a minimally coupled massive Dirac field coupled to gravity, which is the starting point for a quantum theory of spin-1/2 particles in curved spacetimes, can be written
\begin{eqnarray}
{\cal S}_{E-D}[e^{I},\omega^{IJ},\psi]&=&\int\mathcal{L}_P+\mathcal{L}_D=\int \kappa \epsilon_{IJKL}e^I\wedge e^J\wedge (R^{KL}-\frac{\Lambda}{6}e^K\wedge e^L)\nonumber \\&+&\epsilon_{IJKL}(e^I\wedge e^J\wedge e^K\wedge \bar{\psi}\gamma^L D\psi-me^I\wedge e^J\wedge e^K\wedge e^L\bar{\psi}\psi).\label{EinsteinDirac}
\end{eqnarray}
where $R^{IJ}\equiv d\omega^{IJ}+\omega^I_{\ph IK}\wedge \omega^{KJ}$ is the Riemannian curvature two-form and $\Lambda$ is the cosmological constant. The remainder of the paper rests heavily on the calculus of forms. In order to increase readability among tensor-minded physicists we have included several appendices with the necessary techniques and tools of exterior calculus. For example, in Appendix \ref{tensorformtrans} we recall how to translate between the the Palatini action ${\cal S}_P=\int \epsilon_{IJKL}e^I\wedge e^J\wedge R^{KL}$ written in terms of forms and the usual Einstein-Hilbert action ${\cal S}_{EH}=\int d^{4}x\sqrt{-g}R$ by constructing the dual tensor density defined in Appendix \ref{duality}.

The simple action (\ref{EinsteinDirac}) (which is {\em polynomial} in the basic variables) leads in general to non-Riemannian spacetime geometries. Specifically, in the case of non-vanishing spin-density ${\cal J}_{IJ}$, defined by $\delta_\omega\int {\cal L}_D\equiv\int \delta\omega^{IJ}\wedge {\cal J}_{IJ}$, implies non-zero torsion $T^I\equiv de^I+\omega^I_{\ph IJ}\wedge e^J\neq0$ thus violating the zero-torsion condition of Riemannian geometry. However, we note that the Dirac spinor, contrary to the Maxwell field for example, does not represent any classical field observed in nature. Rather it is only its quantized version that corresponds to fermionic matter. Nevertheless, it is expected that the effects of torsion as predicted by a suitable phenomenological theory including spin density are going to be too small to be measured currently \cite{Mao:2006bb} and so we may regard the gravitational part of (\ref{EinsteinDirac}) as a theory having the same experimental support as General Relativity and thus treat it as a legitimate theory of gravity. 

\begin{figure}[!h]
\centering
\includegraphics[width=10cm]{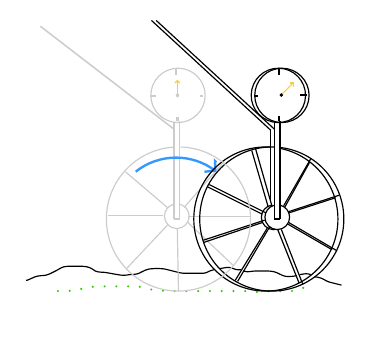} 
\caption{A traditional waywiser, depicted rolling along on a two-dimensional surface. A mechanism converts the rolling of the wheel into a measure of distance traversed along the dotted path, as depicted by the changing orientation of the orange arrow.}
\label{waywiser}
\end{figure}

The method of using the pair of one-forms $\{e^I,\omega^{IJ}\}$ to represent the gravitational field is old and due to \'{E}lie Cartan \cite{cartane,SharpeCartan,Wise:2006sm}. This method has its roots in Cartan's original conception of differential geometry based on symmetric spaces called {\em model spaces} and {\em rolling connections} \cite{SharpeCartan}. The first aim of this paper is to present Cartan geometry as the mathematics of {\em idealized waywisers}. Waywisers were traditionally used to measuring distances between various places, see Fig \ref{waywiser}. The traditional waywiser device is simply a rotating wheel and a `clock' recording how much the wheel has turned. In this way an approximation of the distance covered is obtained. In a more abstract sense, this device is something that can roll along a path on some surface and in doing so yield information about the geometry of the surface (in this case the distance). We will show that a generalization of this device, here denoted an \emph{idealized waywiser}, is capable of probing not just distances but also the nature of the curvature of a surface via its rolling along paths. It shall be seen that the mathematics of this is indeed that of Cartan geometry. In order to put emphasis on the implicit underlying geometric picture in terms of idealized waywisers we shall refer it as \emph{Cartan waywiser geometry}).

The article is organized as follows:  In Section \ref{waywisermath} we develop the mathematical theory of idealized waywisers. In order to facilitate visualization and build intuition, we first restrict attention to the case of two-dimensional manifolds embedded in a three-dimensional space. It is shown that all the basic mathematical objects of Riemannian geometry is recoverable from the mathematical objects that describe the idealized waywiser, the so-called {\em waywiser variables}. Furthermore,  it is shown that torsion and `metric compatibility' admit a simple interpretation in terms of the behaviour of the idealized waywiser. The notion of waywisers and the manner in which they probe geometry is immediately generalizable to manifolds of higher dimension. 
In Section \ref{higherd} we discuss the generalization of Cartan waywiser geometry to the physically important case of four dimensional spacetime manifolds.\footnote{We note that Harvey Brown's book ``Physical Relativity" \cite{BrownPhysicalRelativity} on the foundations on special relativity, from our perspective quite appropriately, depicts a traditional waywiser on its cover!} In Section \ref{standnotation} we clarify the relationship between the waywiser variables and the variables aforementioned variables $e^{I}$ and $\omega^{IJ}$. In Section \ref{actions} we implement these ideas by formulating action principles for gravitation for which the `gravitational field' is characterized entirely by waywiser variables. It is found that vacuum General Relativity may, by different mechanisms, be recovered for both constrained and unconstrained variation of the waywiser variables. Finally in Section \ref{conclusions} we present our conclusions and suggest areas for further exploration.

\section{Introducing Cartan waywiser geometry}\label{waywisermath}
In this section we shall develop the mathematics of idealized waywisers. In this conception of differential geometry both metric and affine connection are {\em derived} concepts and constructed from the more basic waywiser variables whose geometric interpretation is rather straightforward. Let us see how this works. 
\subsection{The mathematics of idealized waywisers}
\label{whatarewaywisers}
Just as in the case of Riemannian geometry it is helpful for the sake of intuition to first invoke an embedding space. Consider then a two-dimensional surface embedded in a three-dimensional Euclidean space and some choice of coordinates $x^a$, $a=1,2$. One may imagine `paths' $x^{a}(\lambda)$ on this surface. We define a waywiser as a device which one may attempt to `roll' along a path
$x^{a}(\lambda)$ and in doing so yield information about the geometry of the surface. The amount of information that may be obtained will depend on the particular nature of the waywiser. The traditional waywiser depicted in Fig. \ref{waywiser} is suitable for measuring physical distances along paths $x^{a}(\lambda)$ on certain surfaces but is otherwise limited by the requirement that it may only roll along any path along the direction tangent to its wheel. A more general notion of a rolling object is a sphere of radius $\ell$. For example, one may imagine a process of rolling such a sphere around a closed path ${\cal C}$. Upon returning the sphere may differ from its original, starting state by an arbitrary rotation, i.e. an $SO(3)$ transformation, which of course is a more general transformation than a traditional waywiser is capable of whilst staying in contact with the surface. 

\begin{figure}[!h]
\centering
\includegraphics[width=10cm]{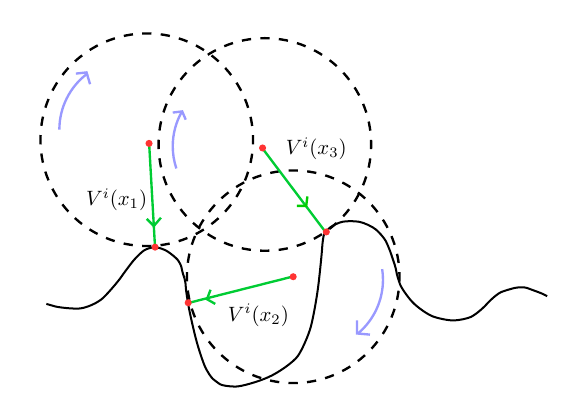} 
\caption {A figure demonstrating that the contact point between the ideal waywiser and manifold is the \emph{only} possible point of contact, and so the ideal waywiser at a given point is `invisible' to all other points. As such, the rolling of the ideal waywiser from $x_{1}\rightarrow x_{2} \rightarrow x_{3}$ is unhindered by features on the surface.}
\label{ghostwaywiser}
\end{figure}

We shall be concerned with what we call {\em idealized waywisers} with symmetric spaces as representing the `wheels'. These are `Platonic' creations of the mind where all irrelevant features, inherent in their material incarnations, have been stripped and abstracted away. For example, no features in the embedded surface may obstruct or hinder the rolling of the idealized waywiser, see Figure \ref{ghostwaywiser}.

The first feature of an idealized waywiser is that it has a contact point between itself and the two-dimensional surface being probed. Such a point of contact is itself a point on the sphere. It is then convenient to represent the contact point by a {\em contact vector} $V^i$ satisfying $V^iV^j\delta_{ij}=\ell^2$ where $\delta_{ij}=diag(1,1,1)$. The Latin index $i=1,2,3$ of the contact vector $V^i$ refers to the three-dimensional Euclidean space. 

Picture now a sphere on top of all the points of the two-dimensional surface. For each sphere we have a contact point which is represented by a vector $V^i$. We note that the contact vector only depends on how the surface is embedded in the three-dimensional Euclidean space and is therefore the same regardless how the waywiser got there. In fact, the contact vector is always normal to the embedded surface. Thus, it is then appropriate to introduce a {\em field} of contact vectors $V^i(x)$ for all the points on the surface. The contact vector $V^i(x)$ at some point $x^a$ we visualize as having its origin in the center of the sphere at the same point $x^a$.

The second feature of the ideal waywiser is a prescription for how the sphere is rotated when rolled from one point to another along some path. Since it is a sphere the transformation group is $SO(3)$. Thus, the rolling of the waywiser corresponds to a succession of infinitesimal $SO(3)$ transformations. Mathematically these infinitesimal transformations can be specified by a connection $A_{a\ph ij}^{\ph ai}$ with values in the Lie algebra of $SO(3)$.\footnote{By the term `$\mathfrak{so}(3)$-valued' is meant that the connection one-form $A_{a\ph ij}^{\ph ai}$, seen as a matrix $(A_a)^i_{\ph ij}$, is a linear combination $(A_a)^i_{\ph ij}=A_a^{\alpha}(S_\alpha)^i_{\ph ij}$ of matrices $(S_\alpha)^i_{\ph ij}$ which satisfies the commutation relations $[S_\alpha,S_\beta]=2i\epsilon_{\alpha\beta}^{\phantom{\alpha\beta}\gamma}S_\gamma$ of the Lie-algebra $\mathfrak{so}(3)$.} By feeding this connection an infinitesimal displacement $\delta x^a$ we obtain an infinitesimal rotation $\delta \Omega^i_j=\delta^i_j-\delta x^aA_{a\ph ij}^{\ph ai}$ \footnote{The minus sign in front of the connection is of course pure convention.}. This infinitesimal rotation characterizes mathematically the infinitesimal `response' of the idealized waywiser and how the point of contact consequently is altered.

We will show in the following section that the notion of ideal waywiser, realized via the fields $\{V^{i},A^{ij}\}$, is sufficient to recover the familiar tensors of Riemannian geometry. More  specifically, all objects in differential geometry can be understood as ways of characterizing the {\em change in the contact point} when the idealized waywiser is rolled.
\begin{figure}[!h]
\centering
\includegraphics[width=10cm]{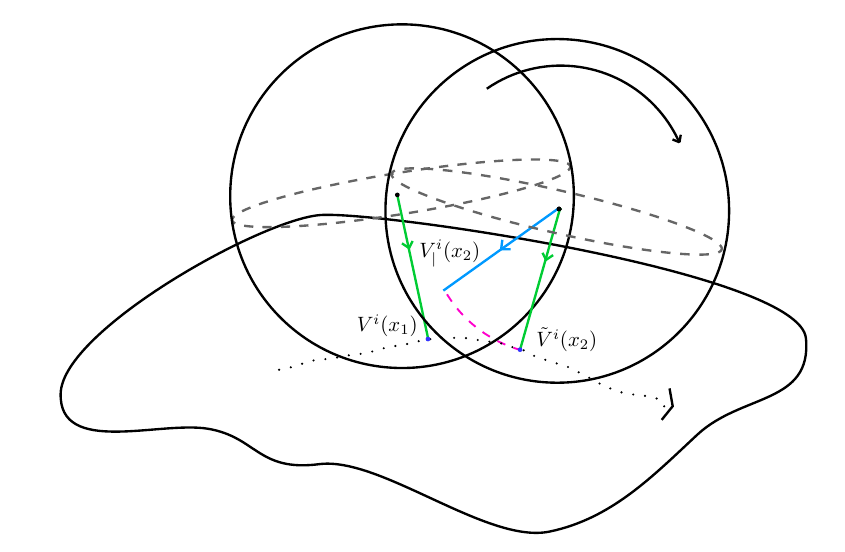}
\caption{The figure illustrates how the `wheel' of the ideal waywiser is rotated when rolled on the surface from point $x_1$ to $x_2$. The contact vectors $V^i(x_1)$ and $V^i(x_2)$ at $x_1$ and $x_2$ respectively can be visualized as having their origins (black dots) in the center of the corresponding sphere, pointing towards the point of contact (the blue dots) between the sphere and the two-dimensional surface. The figure also illustrates how the contact point $V^i(x_1)$ at $x_1$ is `rolled' to $x_2$ yielding $V_|^i(x_2)$ (light blue line). The distance between $x_1$ and $x_2$ is identified as the difference between the rolled $V^i_|(x_2)$ and the contact point $V^i(x_2)$ at $x_2$, i.e. $ds^2=\delta x^a\delta x^bD_aV^iD_bV^j\delta_{ij}$.}
\label{cartangeometry}
\end{figure}
\subsection{Constructing the metric tensor and affine connection}\label{contructmetricaffine}
Let us now determine the distance between two neighboring points $x_1^a$ and $x_2^a$ on the surface. In our mind's eye we now picture an idealized waywiser at $x_1$. Before that ball is rolled we imagine a stick of length $\ell$ attached to the ball, with one end in the center of the ball and the other at the contact point $V^i(x_1)$. We denote this `stick-vector' $V^i_|$ which per definition coincides with the contact vector at $x_1$, i.e. $V^i_|(x_1)=V^i(x_1)$. Next we roll the ball in the direction $\delta x^a=x_2^a-x_1^a$ and put it to rest at $x_2^a$. Rolling the `stick-vector' is mathematically understood as a succession of infinitesimal $SO(3)$ transformations $\delta \Omega^i_j=\delta^i_j-\delta x^aA_{a\ph ij}^{\ph ai}$ acting on $V^i_|$. Thus we have

\begin{eqnarray}
V^i_|(x_2)=\delta \Omega^i_jV^j_|(x_1)=(\delta^i_j-\delta x^aA_{a\ph ij}^{\ph ai})V^j_|(x_1)=V^i(x_1)-\delta x^aA_{a\ph ij}^{\ph ai}V^j(x_1)
\end{eqnarray}
where $A_{a\ph ij}^{\ph ai}V^j$ is the $\mathfrak{so}(3)$-valued one-form dictating how much the ball has rotated and which was introduced in the previous section. Next, we can compare the rolled `stick-vector' $V^i_|(x_2)$ with the contact vector $V^i(x_2)$ at $x_2$ and compute the difference $\delta V^i\equiv V^i(x_2)-V^i_|(x_2)$:
\begin{eqnarray}
\delta V^i&\equiv& V^i(x_2)-V^i_|(x_2)=V^i(x_2)-(V^i(x_1)-\delta x^a A_{a\ph ij}^{\ph ai}V^j(x_1))=\delta x^{a}\partial_aV^i+\delta x^a A_{a\ph ij}^{\ph ai}V^j(x_1)\nonumber\\
&\equiv& \delta x^aD_a V^i
\end{eqnarray}
where we have introduced the gauge covariant derivative $D_a V^i\equiv\partial_aV^i+A_{a\ph ij}^{\ph ai}V^j$. The difference $\delta V^i$ represents the change in contact point. We note that because the contact vector satisfies $V^2=\ell^2$, we have $\delta_{ij}V^iDV^j=0$ and the object $\delta x^aD_aV^i$ therefore has no normal component and belongs to the tangent space of the surface at $x_1$. We now identify the distance $ds$ between the two points $x_1$ and $x_2$ as the Euclidean norm of the difference $\delta V^i$, or equivalently
\begin{eqnarray}
ds^2=\delta_{ij}\delta V^i \delta V^j=\delta x^a\delta x^b\delta_{ij}D_aV^i D_bV^j
\end{eqnarray}
The metric tensor $g_{ab}$, encoding all information about distances of the surface, can then be defined as
\begin{eqnarray}\label{metricdef}
g_{ab}=\delta_{ij}D_aV^i D_bV^j.
\end{eqnarray}
We have now understood how distances, and in particular the metric tensor, can be recovered from the waywiser variables $\{V^i,A^{ij}\}$. In particular, we see that the metric directly corresponds to the change of contact point when the waywiser is rolled. However, the metric tensor cannot tell us how to parallel transport tangent vectors, $u^a$ say, along the surface, something which is encoded in the affine connection $\Gamma^c_{ab}$. Nevertheless, also this mathematical object can easily be constructed from the waywiser variables and is related to the {\em rate} of change of the contact vector. More specifically, the object $D_aD_bV^i=\partial_a D_bV^i+A_{a\ph ij}^{\ph ai}D_bV^j$ contains components both normal and tangential to the embedded surface. It is easily checked that the normal component is the metric. It is in the tangential part that we can identify an affine connection $\Gamma^a_{bc}$. Thus we define
\begin{eqnarray}\label{affinedef}
P^i_{\ph ij}D_aD_b V^j\equiv\Gamma^c_{ab}D_cV^i.
\end{eqnarray}
where and $P^i_{\ph ij}\equiv\delta^i_j-\frac{1}{\ell^2}V^iV_j$ is a projector. We note that, as should be the case, both the left- and right-hand side do not transform as tensors. We see that the affine connection can be recovered from the waywiser variables $\{V^i,A^{ij}\}$ and consequently all the information of how to parallel transport tangent vectors. In addition, we recover the covariant derivative $\nabla_a$ acting on tensors from which we can construct the Riemann curvature tensor $R^a_{\ph abcd}$. We see that all the objects of Riemannian geometry can be extracted, if needed, from  the waywiser variables. 

It is quite pleasing to see that both metric and affine connection, which play two distinct mathematical roles in Riemannian geometry, can be constructed from the more primary variables $\{V^i,A^{ij}\}$ which themselves admit a crisp geometric interpretation in terms of idealized waywisers. In a sense we can say that going from Riemannian geometry to Cartan waywiser geometry is an instance of unification since the metric tensor and affine connection, whose roles are conceptually and mathematically distinct, are seen merely as two aspects of the response of idealized mathematical waywiser when rolled. Indeed, all of differential geometry is now understood merely as different ways of characterizing the change of contact point that the waywiser undergoes when rolled. Therefore, it does not seem too preposterous to say that Cartan waywiser geometry, simply being the mathematics of easily visualized waywisers, is both conceptually and mathematically simpler than Riemannian geometry.

\subsection{Abstract Cartan waywiser geometries}
We can now forget about the embedding space which only served to facilitate visualization and helping intuition along. The situation is not different from Riemannian geometry where embedding spaces are invoked to facilitate visualization. The mathematical representation of an abstract Cartan waywiser geometry is simply the pair $\{V^i,A^{ij}\}$ and no reference to an embedding space is required. From a mathematical point of view we see that we are dealing with a fiber bundle structure where the base space is the manifold, the fiber the sphere, and the structure group $SO(3)$. However, it is easier to work with a three-dimensional vector $\mathbb{R}^3$ space as the fiber instead of the two-dimensional sphere $S^2$. The contact point is then represented by a contact vector $V^i\in\mathbb{R}^3$ subject to the constraint $V^2=\ell^2$ and the variable $A_{\ph ij}^{i}$ is a gauge connection on that vector bundle. It should be clear that, although it is helpful to imagine embedding spaces, we can understand Cartan geometry abstractly in terms of this fiber bundle structure.
\subsection{Metric compatibility and torsion}\label{metricitorsion}
The space of all possible pairs $\{g_{ab},\Gamma^c_{ab}\}$ can be `coordinatized' by the non-metricity tensor $Q_{cab}\equiv\nabla_cg_{ab}$ and the torsion tensor $T^c_{ab}\equiv\Gamma^c_{ab}-\Gamma^c_{ba}$ \cite{SchoutenRic}. Let us now consider the space of pairs $\{g_{ab},\Gamma^c_{ab}\}$ that can be be generated by the waywiser variables $\{V^i,A^{ij}\}$. We turn first to metricity. Given the expressions (\ref{metricdef}--\ref{affinedef}) for the metric tensor and affine connection we can compute
\begin{eqnarray}
\nabla_{c}g_{ab}&\equiv&\partial_c g_{ab}-\Gamma^d_{ca}g_{db}-\Gamma^d_{cb}g_{ad}=\partial_c g_{ab}-(\Gamma^d_{ca}D_dV^iD_bV^j+\Gamma^d_{cb}D_aV^iD_dV^j)\delta_{ij}\nonumber\\
&=&\partial_{c}g_{ab}-(P^i_{\ph ik}D_cD_aV^kD_bV_i+P^j_{\ph jk}D_aV_iD_cD_bV^k)\delta_{ij}=\partial_{c}g_{ab}-D_c(D_aV^iD_bV^j)\delta_{ij}\nonumber\\
&=&\partial_{c}g_{ab}-D_c(D_aV^iD_bV^j\delta_{ij})=\partial_{c}g_{ab}-\partial_{c}g_{ab}\equiv0\nonumber
\end{eqnarray}
where we made use of the fact that $P^i_{\ph ij}D_aV^j=D_aV^i$ and that the gauge group is the orthogonal group $SO(3)$ so that $D_a\delta_{ij}=0$. Thus we see that metric compatibility $\nabla_cg_{ab}=0$ is deduced and not postulated. It is a consequence of the fact that we are dealing with rolling a sphere with symmetry group $SO(3)$ whose gauge connection satisfies $A^{ij}=-A^{ji}$.

Let us now turn to torsion and its geometrical interpretation within our approach. Note that there is no guarantee that the affine connection as defined by \eqref{affinedef} is symmetric. Indeed, its antisymmetric part is given by
\begin{eqnarray}\label{torsiondef}
F^{\ph{ab}i}_{ab\ph ij}V^j=P^i_{\ph ik}F^{\ph{ab}k}_{ab\ph kj}V^j\equiv P^i_{\ph ij}[D_a,D_b]V^j=(\Gamma^a_{bc}-\Gamma^a_{cb})D_aV^i\equiv T^a_{bc}D_aV^i
\end{eqnarray}
where $T^a_{bc}$ is the torsion tensor. In the Cartan waywiser geometry, torsion has a very simple geometric interpretation. The left-hand-side of \eqref{torsiondef} represents mathematically how much the contact vector has changed when parallel transported around an {\em infinitesimal} closed loop. We see that torsion is merely a particular aspect of the $SO(3)$ curvature  $F^{ij}$. In fact, Riemannian curvature and torsion are aspects of the same thing: it is the non-integrability of the $SO(3)$ connection. This unification of torsion and Riemannian curvature is very pleasing and will suggest a natural modification of the gravitational field equations as we shall see in Section \ref{holst}.
\section{Waywisers for General Relativity}
\label{higherd}
Now that we have gained some intuition about Cartan geometry and its geometric interpretation in terms of idealized waywisers, we turn to General Relativity. To accommodate spacetime geometries and relativistic theories we must adapt the above waywiser formalism accordingly. From a mathematical point of view the obvious change to make is to make use of symmetric spacetimes, rather than spaces, as idealized waywiser `wheels'. In the literature the symmetric spacetimes representing idealized relativistic waywiser wheels go by the name {\em model spaces} or {\em model spacetimes}. We shall from now on use those terms interchangeably.

In this article we will focus on two choices of model spacetimes: the De Sitter and anti-De Sitter spacetimes. We could also use a flat Minkowski spacetime as model spacetime. But this choice of model spacetime requires a slightly different mathematical representation \cite{Gronwald:1995em} of the contact point and we will not discuss that option in this paper \cite{Gronwald:1995em,Wise:2006sm}.
\subsection{De Sitter spacetime as model spacetime}
As a first mathematical realization of the idealized `relativistic wheel', i.e. model spacetime, we consider the De Sitter spacetime defined by 
\begin{eqnarray}
-t^2+x^2+y^2+z^2+w^2=\ell^2
\end{eqnarray}
which has the symmetry group $SO(1,4)$, and a spacelike contact vector $V^A$ satisfying $V^AV^B\eta_{AB}=\ell^2$, $\eta_{AB}=diag(-1,1,1,1,1)$, where $A=0,\dots 4$. The spacetime waywiser is then represented by the pair $\{V^A(x),A^{AB}(x)\}$, where $A^{AB}=A_\mu^{\ph \mu AB}dx^\mu$ is a $\mathfrak{so}(1,4)$-valued one-form, where $\mu=0,\dots 3$. The subgroup of transformations that leave the components of the spacelike contact vector $V^A$ invariant is just the Lorentz group $SO(1,3)$.
\subsection{Anti-De Sitter spacetime as model spacetime}\label{antiwheel}
Our second choice for model spacetime is the anti-De Sitter spacetime defined by
\begin{eqnarray}
-t^2+x^2+y^2+z^2-w^2=-\ell^2
\end{eqnarray}
which has the symmetry group $SO(2,3)$. The contact vector $V^A$ is timelike, rather than spacelike, and satisfies $V^AV^B\eta_{AB}=-\ell^2$ with $\eta_{AB}=diag(-1,1,1,1,-1)$. The pair $\{V^A(x),A^{AB}(x)\}$ denotes the spacetime waywiser variables where $A^{AB}$ is a $\mathfrak{so}(2,3)$-valued one-form. The subgroup which leaves the components of this timelike contact vector $V^A$ invariant is again the Lorentz group $SO(1,3)$.

In the following we will consider both model spacetimes simultaneously and so shall not make a notational distinction between the two $\eta_{AB}$'s, corresponding to De Sitter and anti-De Sitter model spacetimes. As for the signs, e.g. $V^2=\mp1$, that will appear from now on, we understand the upper sign as referring to the anti-De Sitter model spacetime and the lower to the De Sitter one. 
\section{Relation to standard notation}\label{standnotation}
The formalism and choice of mathematical variables in this article serves to highlight the idea that Cartan geometry is simply the mathematics of idealized waywisers. The Cartan waywiser formalism has inbuilt $SO(p,q)$ symmetry (with $(p,q)=(1,4)$ or $(2,3)$) and we can make use of that gauge redundancy to fix the contact vector to be everywhere equal to $V^A(x)\overset{*}{=}\ell \delta^A_4$ . For such a gauge choice we make contact with the more standard variables used in Cartan geometry. We can identify the co-tetrad $e^I$ and spin connection $\omega^{IJ}$ in the following way:
\begin{eqnarray}\label{standardrel1}
e^A\equiv DV^A=dV^A+A^A_{\ph AB}V^B\overset{*}{=}\mp\ell A^{A4}=(e^I,0) \qquad \omega^{AB}\equiv h^A_{\ph AC}h^B_{\ph BD}A^{CD}\overset{*}{=}\left(\begin{array}{cc}\omega^{IJ}&0\\0&0\end{array}\right)
\end{eqnarray}
where $h^A_{\ph AB}\equiv \delta^A_B-\frac{V^AV_B}{V^2}$ is a projector. We note that while the definition of the co-tetrad $e^A$ includes a gauge covariant exterior derivative, this is not the case for the spin-connection $\omega^{AB}$. This signals a significant mathematical difference between the two objects. In particular, while the spin connection $\omega^{AB}$ transforms inhomogeneously under a $SO(p,q)$ gauge transformation, the same is not true for the co-tetrad $e^A$. For this reason the co-tetrad $e^I$ cannot be thought of as a gauge connection in this context.\footnote{We contrast our approach to Poincar\'e gauge theory \cite{Hehl:1994ue} in which the co-tetrad is conceptualized as a gauge connection with respect to local translations.} Specifically, the co-tetrad should not be thought of as a gauge connection related to the `translational' symmetry of the De Sitter or anti-De Sitter model spacetimes\footnote{A more accurate term is {\em transvections} \cite{Randono:2010cq}.}. Rather, the co-tetrad is best understood as the quantifying the change of contact point when the idealized waywiser wheel is rolled; something which is not a gauge quantity. 

The $SO(p,q)$ curvature two-form $F^{AB}$ can be split into a projected part $h^A_{\ph AC}h^B_{\ph BD}F^{CD}$ and a normal part $F^{AB}V_B$. The projected curvature two-form is the Riemannian curvature $R^{IJ}$ two-form but `corrected' by the curvature of the model spacetime, and the normal part is simply the torsion, i.e. we have 
\begin{eqnarray}\label{standardrel2}
h^A_{\ph AC}h^B_{\ph BD}F^{CD}&\overset{*}{=}&\left(\begin{array}{cc}dA^{IJ}+A^I_{\ph IC}\wedge A^{CJ}&0\\0&0\end{array}\right)=\left(\begin{array}{cc}R^{IJ}\pm\frac{1}{\ell^2}e^I\wedge e^J&0\\0&0\end{array}\right)\\ T^A&\equiv& F^A_{\ph AB}V^B\overset{*}{=}(\mp\ell F^{I4},0)=(T^I,0)
\end{eqnarray}
where the sign as prescribed in section \ref{antiwheel}.

From the perspective developed in Section \ref{waywisermath} we see that the gauge fixing, although useful, obscures the underlying geometric picture in terms idealized waywisers which are mathematically represented by both contact point $V^A$ and rolling connection $A^{AB}$. If we resist the temptation of immediately gauge $V^A$ `out of existence', mathematical and conceptual clarity is increased. We now proceed to see under what circumstances gravitation may be understood as a theory of Cartan waywiser geometry.
\section{Action principles for gravity}
\label{actions}
The Einstein-Hilbert action ${\cal S}_{EH}=\int \sqrt{-g} g^{\mu\nu}R_{\mu\nu}d^4x$ is a rather complicated action. It is manifestly non-polynomial in its basic dynamical variable $g_{\mu\nu}$ (since it involves the square root $\sqrt{-g}$ of the metric determinant $g=det\ g_{\mu\nu}$) as well as the inverse metric $g^{\mu\nu}$. The action is further complicated by the fact that it contains second order partial derivatives with respect to the metric tensor. This makes it necessary to add, in the case of non-compact spaces, a compensating non-local boundary term in order to ensure that the Einstein-Hilbert action is indeed extremized whenever the field equations are satisfied \cite{Wald:1984rg}.

On the other hand, the natural actions for General Relativity using the waywiser variables $\{V^A,A^{AB}\}$, are {\em polynomial} in the basic waywiser variables, and are, from a mathematical point of view, the simplest actions possible. This is due to the fact that the waywiser variables are all forms; $V^A$ is a zero-form and $A^{AB}$ a one-form. Since an action is per definition an integration over a four-form, the construction of the simplest actions possible in Cartan waywiser geometry is just an exercise in `wedging' together the various forms we can construct from the waywiser variables.\footnote{Non-polynomial actions for General Relativity based on gauge connections can be considered \cite{Krasnov:2011pp,Krasnov:2012pd} but we shall restrict attention to polynomial actions.} Building an action is very much like playing with Lego \cite{Lego}: You only have but a few basic pieces (the forms) and the only task is to find out how to fit the pieces together to create four-forms. 

Before we start playing with waywiser forms, we note that where are two distinct approaches to obtain viable actions for gravity which are equivalent, at least in the vacuum case. These are:
\begin{itemize}
\item {\bf Non-dynamical:} $V^A$ is regarded as a non-dynamical {\em \`a priori} postulated variable, also called an absolute object \cite{AndersonRelativity,WestmanSonego2007b}. We simply pick some contact field $V^A(x)$ subject to the only constraint $\eta_{AB}V^{A}V^{B}=\pm \ell^2$. Neither are equations of motion given for the contact vector $V^A$ nor is it necessary. Diffeomorphism invariance is broken except in the special $SO(p,q)$ gauge in which $V^A=\ell\delta^A_4$. 
\item {\bf Dynamical:} $V^A$ is regarded as a dynamical variable on par with $A^{AB}$ which have its own equations of motion and should be varied with respect to in an action principle. This requires a non-standard choice of action in order to ensure consistency with the standard Einstein vacuum field equations. This formulation is manifestly diffeomorphism invariant.
\end{itemize}
In the following we shall pursue both views. The following sections will make heavy use of the variational calculus of forms. For an exposition of all necessary ideas and techniques of the variational calculus of forms we point to Appendix \ref{variationalforms}.
\subsection{A class of polynomial actions for gravity}\label{secgenaction}
Let us then contemplate what kind of Lagrangian polynomial four-forms $\mathcal{L}$ may be constructed. To do that we should first list the basic building blocks we have at our disposal.
\begin{itemize}
\item the waywiser variables $\{V^A,A^{AB}\}$ from which the gauge covariant objects $F^{AB}$ and the one-form $DV^{A}$ can be constructed
\item the `internal' Minkowski metric $\eta_{AB}$ and Levi-Civita symbol $\epsilon_{ABCDE}$ associated with the orthogonal groups $SO(1,4)$ or $SO(2,3)$.
\end{itemize}
The most general polynomial gravitational action that can be constructed is
\begin{eqnarray}\label{genaction}
{\cal S}_{g}=\int a_{ABCD} F^{AB}\wedge F^{CD} &+& b_{ABCD} DV^A\wedge DV^B\wedge F^{CD}\nonumber\\
&+&c_{ABCD} DV^A\wedge DV^B\wedge DV^C\wedge DV^D
\label{actione}
\end{eqnarray}
where 
\begin{eqnarray}
a_{ABCD} &=& a_{1}\epsilon_{ABCDE}V^{E}+a_{2} V_{A}V_{C}\eta_{BD} +a_{3} \eta_{AC}\eta_{BD} \\
b_{ABCD} &=& b_{1}  \epsilon_{ABCDE}V^{E}+b_{2} V_{A}V_{C}\eta_{BD}+b_{3}\eta_{AC}\eta_{BD}\\
c_{ABCD} &=&  c_{1}\epsilon_{ABCDE}V^{E}
\end{eqnarray}
In general the quantities $a_{i},b_{i},c_{i}$ may depend on the scalar $V^{2}=V_{E}V^{E}$. We shall however restrict ourself to the case where they are just constants. Given this assumption, we note from (\ref{top1})  that the $a_{3}$ term is topological and we see from \eqref{nyny} that the the $a_{2}$ and $b_{3}$ terms are topologically equivalent; therefore in this case only five of the $a_{i},b_{i},c_{i}$ independently contribute to the equations of motion, namely $a_1,a_2,b_1,b_2$, and  $c_1$.
\subsection{The contact vector as non-dynamical absolute object}\label{nondynapproach}
In the non-dynamical view we regard the contact vector as postulated and not subject to equations of motion. In a generic $SO(p,q)$ gauge choice the field $V^A(x)$ breaks diffeomorphism invariance since $V^A(x)$ depends explicitly on the coordinate $x^\mu$. The situation is similar to a Klein-Gordon field in flat spacetime. The action ${\cal S}_{KG}$ contains a non-dynamical and {\em \`a priori} postulated symmetric tensor $\eta_{\mu\nu}$, subject to the requirement of being flat and having signature $+2$. We do not require any equations of motion for $\eta_{\mu\nu}$ and the action ${\cal S}_{KG}$ should not be varied with respect to $\eta_{\mu\nu}$ since that would only yield nonsensical equations.

However, while the Klein-Gordon theory is not diffeomorphism invariant, the diffeomorphism invariance of the waywiser action is restored in the particular $SO(p,q)$ gauge where $V^A(x)\overset{*}{=}\ell \delta^A_4$. There is thus an curious interplay between diffeomorphism and $SO(p,q)$ gauge invariance.
\subsubsection{The MacDowell-Mansouri action}
Let us now consider  actions appropriate within the non-dynamical view. The simplest action we can write down is known as the MacDowell-Mansouri action \cite{MacDowell:1977jt}
\begin{eqnarray}\label{MMactioneq}
{\cal S}_{MM}=\int\mathcal{L}_{MM}=\int\kappa\epsilon_{ABCDE}V^E F^{AB}\wedge F^{CD} \label{mmaction}
\end{eqnarray}
and corresponds to only having $a_1$ non-zero in the general action \eqref{genaction}. The equations of motion are obtained by varying {\em only} with respect to the connection $A^{AB}$ and not with respect to the contact vector $V^A$ which is here treated as a non-dynamical absolute object. In Appendix \ref{MMaction} the variation is done in pedagogical detail and yields:
\begin{eqnarray}\label{MMeqs}
\epsilon_{ABCDE} DV^E\wedge F^{CD}=0.
\end{eqnarray}
These polynomial equations, which are written in a rather succinct form, are equivalent to Einsteins field equations. To see this we impose the gauge choice $V^A=\ell \delta^A_4$, make use of the relations \eqref{standardrel1} and \eqref{standardrel2}, put $A=4,B=I$ and $A=I,B=J$ in equation \eqref{MMeqs} which yields respectively the two equations
\begin{eqnarray}
\epsilon_{IJKL} e^L\wedge (R^{JK}\pm\frac{1}{\ell^2}e^J\wedge e^K)&=&0\label{riemMM}\\
\frac{1}{\ell}\epsilon_{IJKL} e^L\wedge T^K&=&0\label{torsionMM}
\end{eqnarray}
These equations may look unfamiliar but are nothing but the Einstein field equations with cosmological constant and the torsion-free condition. In Appendix \ref{standardform} we show how in pedagogical detail how the equations of motion \eqref{MMeqs} can be rewritten in tensor notation as
\begin{eqnarray}
R_\mu^{\phantom{\mu}\nu}-\frac{1}{2}\delta_\mu^{\phantom{\mu}\nu} R+\frac{6}{\ell^2}\delta_\mu^{\phantom{\mu}\nu}=0 \qquad T^\rho_{\mu\nu}=0.
\end{eqnarray} 
Although the MacDowell-Mansouri action is the simplest possible action we can write down, it does not appear natural from a Cartan waywiser geometry point of view. As we noted in section \ref{metricitorsion}, torsion  $T^A=F^A_{\ph AB}V^B$ in Cartan waywiser geometry is merely a particular aspect of the $SO(p,q)$ curvature and as such we would expect it to appear in a symmetric fashion in a gravitational action. However, the MacDowell-Mansouri action (\ref{MMactioneq}) does not contain torsion since any normal component $F^A_{\ph AB}V^B$ of the curvature two-form is projected out by the factor $\epsilon_{ABCDE}V^E$ in the action. From a waywiser geometry point of view, there is therefore a strange asymmetry in the MacDowell-Mansouri action.

We further highlight this by fixing the $SO(p,q)$ gauge so that $V^A=\ell\delta^A_4$ in which case the MacDowell-Mansouri action can be rewritten as follows
\begin{eqnarray}
\int\mathcal{L}_{MM}&=&\int\kappa\epsilon_{IJKL}\ell (R^{IJ}\pm\frac{1}{\ell^2}e^I\wedge e^J)\wedge (R^{KL}\pm\frac{1}{\ell^2}e^K\wedge e^L)\\
&=&\pm\int\kappa\epsilon_{IJKL}\ell \left(\frac{2}{\ell^2}e^I\wedge e^J\wedge R^{KL}\pm\frac{1}{\ell^4}e^I\wedge e^J\wedge e^K\wedge e^L\right).
\end{eqnarray}
where the topological term $\epsilon_{IJKL} R^{IJ}\wedge R^{KL}$, known as the Euler four-form, was discarded (see Appendix \ref{topologicalterms}). This action is the standard Palatini action with positive or negative cosmological constant depending on the choice of model spacetime. Again we see that the MacDowell-Mansouri action contains only the $SO(1,3)$ Riemannian curvature and not torsion. 
\subsubsection{The Holst action}\label{holst}
From the point of view of waywiser geometry a more natural-looking action can be obtained by adding an extra term which corresponds to $a_2$-term $DV_A\wedge DV_B\wedge F^{AB}$ defined in Section \ref{secgenaction} and is known in the literature as the {\em Holst term} \cite{Holst:1995pc}. The resulting action is the starting point of loop quantum gravity and related to the Ashtekar formulation of gravity \cite{Ashtekar:1986yd,Thiemann:2007zz}.

The Holst action is
\begin{eqnarray}\label{Holstaction}
{\cal S}_{Holst}=\int\mathcal{L}_{Holst}=\int(\epsilon_{ABCDE}V^E + \beta V_AV_C \eta_{BD}) F^{AB}\wedge F^{CD}.
\end{eqnarray}
In order for the units in the action to work out the dimension of $\beta$ is inverse length. The Holst term is topologically equivalent to the squared torsion term $T^A\wedge T_A$ since their difference is a exterior derivative of the three-form called the Nieh-Yan three-form, see Appendix \ref{topologicalterms}. The Holst action therefore contains both Riemann curvature and torsion and we see that the Holst term have restored the asymmetry between Riemannian curvature and torsion of the MacDowell-Mansouri action. From a Cartan waywiser geometry perspective this is more natural since torsion and Riemannian curvature are merely two aspects of the $SO(p,q)$ curvature.

Since neither the Holst term nor the square torsion term $T^A\wedge T_A$ are topological we cannot simply add them without also changing the equations of motion. However, even though the Holst term changes the equations of motion, the predictions are equivalent to General Relativity when the spin density three-form ${\cal J}_{IJ}$ vanishes. Let us see how that comes about. The equations of motion are as in the MacDowell-Mansouri case obtained by varying only with respect to the connection $A^{AB}$ and not $V^A$. This yields
\begin{eqnarray}
\left(2\epsilon_{ABCDE} DV^E+\beta(DV_A\eta_{BD}V_C-DV_B\eta_{AD}V_C+V_A\eta_{BD}DV_C-V_B\eta_{AD}DV_C)\right)\wedge F^{CD}=0
\end{eqnarray}
Let us now look at these set of equations in the particular gauge $V^A=\ell\delta^A_4$. If we set $A=4,B=I$ and $A=I,B=J$ we respectively obtain the two equations
\begin{eqnarray}
2\epsilon_{IJKL} e^L\wedge (R^{JK}\pm\frac{1}{\ell^2}e^J\wedge e^K)\pm\beta\ell D^{(\omega)}T_I&=&0\label{riemholst}\\
\pm4\epsilon_{IJKL} e^K\wedge T^L+\beta\ell(e_I\wedge T_J-e_J\wedge T_I)&=&0\label{torsionholst}
\end{eqnarray}
where $D^{(\omega)}T^I\equiv dT^I+\omega^I_{\ph IJ}\wedge T^J=R^I_{\phantom{I}J}\wedge e^J$. By taking the `internal dual' of the second equation \eqref{torsionholst}, using the `internal' Levi-Civita symbol $\epsilon_{MN}^{\phantom{MN} IJ}$, we obtain
\begin{eqnarray}
&&\frac{1}{2}\epsilon_{MN}^{\phantom{MN} IJ}\left(\pm4\epsilon_{IJKL} e^K\wedge T^L+\beta\ell(e_I\wedge T_J-e_J\wedge T_I)\right)\nonumber\\
&=&\mp 4(e_M\wedge T_N-e_N\wedge T_M)+\beta\ell(\epsilon_{MNKL} e^K\wedge T^L) = 0 \label{dualeq}
\end{eqnarray}
which looks almost like the original equation \eqref{torsionholst} but with the numerical factor $\beta\ell$ appearing on the other term. This comes about because the two terms are essentially the duals of each other. Solving \eqref{dualeq} yields
\begin{eqnarray}
\pm4( e_I\wedge T_J-e_J\wedge T_I)=\beta\ell\epsilon_{IJKL} e^K\wedge T^L
\end{eqnarray}
which we insert in \eqref{torsionholst} which in turn yields
\begin{eqnarray}
(16+\beta^2\ell^2)\epsilon_{IJKL} e^K\wedge T^L=0.
\end{eqnarray}
If we require the action to be real-valued \footnote{For a discussion of the complex-valued (anti-)self-dual cases in which $\beta=\pm \frac{4i}{\ell}$ see \cite{Thiemann:2007zz}.}, so so also $\beta$, we see that $16+\beta^2\ell^2\neq0$ and we obtain the equation $\epsilon_{IJKL} e^K\wedge T^L=0$ which is the same zero torsion equation \eqref{torsionMM} obtained from the MacDowell-Mansouri action. Thus, we conclude that in the absence of fermionic matter torsion is again zero. After torsion has been removed from \eqref{riemholst} what remains is simply Einstein's vacuum equations. Thus, the Holst action reproduces the Einstein's General Relativity.

It should be stressed that the Holst term {\em does} change the way fermionic matter couples to gravity and by changing the value of $\beta$ we get different behavior of the gravitational field inside spacetime regions with non-zero spin-density. The value of $\beta$ is therefore ultimately an experimental question.

As previously stated, the Holst action is more pleasing than the MacDowell-Mansouri action from a Cartan waywiser point of view. It does not appear natural that only the projected part of the $SO(p,q)$ curvature should appear in the action. After all, torsion is merely a special part of the curvature $F^{AB}$.
\subsection{The contact vector $V^A$ as dynamical field}\label{dynapproach}
We now explore the second approach wherein the contact vector $V^A$ is treated as as just another dynamical field, i.e. we require that the gravitational action is also stationary with respect to small variations of $V^{A}$. By turning the contact vector into a dynamical field we increase the number of field equations by five. It is therefore a possibility that the new field equations impose unreasonable constraints and narrowing the space of solutions accordingly. For example, if we consider the MacDowell-Mansouri action \eqref{MMactioneq} and regard  $V^A$ as a dynamical field we obtain, by varying the action with respect to $V^A$, the five additional field equations $\epsilon_{ABCDE}F^{AB}\wedge F^{CD}=0$. It may be checked that this implies a restriction that the Pontryagin four form $\epsilon_{IJKL}R^{IJ}\wedge R^{KL}$ vanishes. Therefore the $V^{E}$ equations of motion merely restrict the solution space to be smaller than that of General Relativity rather than producing equations for $V^{E}$ itself.

Furthermore, in order for $V^A$ to be interpreted as representing a contact point (see Section \ref{waywisermath}), and to reproduce the Einstein's gravitational theory, it must satisfy $V^2=\mp\ell^2$. Since no restrictions are imposed {\em \`a priori} on the dynamical field $V^A$, the condition $V^2=\mp\ell^2$ must somehow be a consequence of the equations of motion. Of course, this can be achieved by simply adding a Lagrange multiplier to the MacDowell-Mansouri action \eqref{MMactioneq} or Holst action \eqref{Holstaction} \cite{Stelle:1979va,Pagels:1983pq,Pagels:1982tc,Randono:2010ym}:
\begin{eqnarray}
{\cal S}_{\lambda}[\lambda,V^{A}] =  \int \lambda \left(V^{2} \pm 1\right)
\end{eqnarray}
where the sign is determined by the choice of model spacetime as prescribed in section \ref{antiwheel}. Requiring that the action is stationary with respect to small variations of the Lagrange multiplier four-form $\lambda$ then produces the required fixed norm constraint. But this procedure is artificial since rather than enforcing equations of motion of dynamical variables, the equations of motion for $V^{E}$ simply amount to a definition of $\lambda$.

The problem to come up with natural action where $V^A$ is itself a dynamical field was labeled an open problem \cite{Randono:2010cq} and has inspired attempts at providing an action where $V^A$ can be regarded as a dynamical field, see e.g. \cite{Randono:2010ym}. We now show that such an action can be found among the general class of polynomial actions (\ref{actione}): those for which only $b_{1}$ and $c_{1}$ are nonzero. No Lagrange multiplier is necessary and the constancy and sign of $V^2$ are consequences of the dynamical equations. The result applies to vacuum and how to include matter fields we leave as an open problem. In addition, it would be interesting with this result could also be generalized to include the Holst term.
\subsubsection{Equations of motion}
Consider then the action
\begin{eqnarray}\label{bcaction}
{\cal S}_g[V^{A},A^{AB}]=\int b_1 \epsilon_{ABCDE}V^E DV^A\wedge DV^B\wedge F^{CD}+c_1\epsilon_{ABCDE}V^E DV^A\wedge DV^B\wedge DV^C\wedge DV^D
\end{eqnarray}
The equations of motion for \eqref{bcaction} follow from requiring stationarity of the action under small variations of the fields $A^{AB}$ and $V^{A}$ yields:
\begin{eqnarray}
\delta {\cal S}_{g}[V^{A},A^{AB}]&=& \int \left(  \delta A^{AB} \wedge {\cal E}_{AB} + \delta V^{A} {\cal E}_{A}\right)=0
\end{eqnarray}
where it has been assumed that both $\delta A^{AB}$ and $\delta V^{A}$ vanish on the boundary of integration and where we have defined
\begin{eqnarray}
{\cal E}_{AF}&\equiv& 2\epsilon_{[A|BCDE}V_{|F]}V^{E}e^B\wedge\left(b_{1} F^{CD}+2c_{1}e^C\wedge e^D\right) -b_{1} \epsilon_{ABCDF}\left(e^B\wedge e^C+2V^{B}T^{C} \right)\wedge e^{D}\nonumber\\
{\cal E}_E&\equiv& b_{1} \epsilon_{ABCDE}\left(3 e^A\wedge e^B+2V^{A}T^{B}\right)\wedge F^{CD}  +c_{1} \epsilon_{ABCDE}\left(5e^A\wedge e^B+ 12V^{A}T^{B}\right)\wedge e^C\wedge e^D\nonumber
\end{eqnarray}
where we recall that $T^{B}\equiv F^{B}_{\phantom{B}C}V^{C}$ and $e^A\equiv DV^A$. The first equation ${\cal E}_{AF}=0$, obtained by varying the action with respect to $A^{AB}$, is a system of ten three-form equations, whilst the second equation ${\cal E}_E=0$, obtained by varying the action with respect to $V^{A}$, is a system of five four-form equations. 
\subsubsection{Constancy and sign of $V^2$ deduced from equations of motion}
No restriction has been placed so far on the norm of $V^{E}$, so solutions where it is constant and non-vanishing must arise from the equations of motion themselves. We will now show that this is the case. To do this we consider the equations $V^E{\cal E}_E=0$ and $e^A\wedge{\cal E}_{AF}V^F=0$ which after simplification takes the form
\begin{eqnarray}
V^E{\cal E}_E&=& \epsilon_{ABCDE}V^{E} e^A\wedge e^B\wedge\left(3b_{1} F^{CD}+5c_{1} e^C\wedge e^D\right)=0\label{eq1}\\
e^A\wedge{\cal E}_{AF}V^F&=&\epsilon_{ABCDE}V^Ee^A\wedge e^B\wedge\left(b_1 e^C\wedge e^D-V^2(2c_1 e^C\wedge e^D+b_1 F^{CD})\right)=0\label{eq2}
\end{eqnarray}
The first equation \eqref{eq1} can now be used to eliminate the curvature two-form $F^{CD}$ in the second equation \eqref{eq2}. This yields the equation
\begin{eqnarray}
\left(b_1-\frac{c_1}{3}V^2\right)\epsilon_{ABCDE}V^{E}e^A\wedge e^B\wedge e^C\wedge e^D=0 \label{vev}
\end{eqnarray}
and for non-degenerate co-tetrads $e^A$ we deduce that this equation is solved only if
\begin{eqnarray}
V^2=\frac{3b_1}{c_1}=\mp\ell^2.
\end{eqnarray}
Since $b_1$ and $c_1$ are constants we see that $V^2$ is constant. We also note that the sign of $V^2$ is determined by the relative sign of $b_1$ and $c_1$. This means that the dynamical equations also determine choice of the model spacetime. In the case where $b_1$ and $c_1$ have opposite sign we need to use the anti-De Sitter model spacetime and De Sitter spacetime for equal sign.
\subsubsection{Consistency with Einstein's vacuum field equations}
Within the dynamical approach there are five additional field equations associated with the variable $V^A$. It is therefore not clear whether this theory contains all the solutions of Einstein's General Relativity. We shall demonstrate consistency with General Relativity in the case of vacuum and leave the inclusion of matter as an open problem.

If we impose the special gauge in which $V^{A}= \ell \delta^{A}_{4}$ and use the notation of Section \ref{standnotation}, the equations ${\cal E}_{IJ}$ becomes
\begin{eqnarray}
0=b_{1}\ell\epsilon_{IJKL} T^{K} \wedge e^{L} \label{tors}
\end{eqnarray}
which implies that the torsion tensor is zero. Let us now study the remaining equations ${\cal E}_{I}$ and see if they in any way restrict the solution space of General Relativity. After simplification we get
\begin{eqnarray}
2b_1\epsilon_{IJKL}T^J\wedge\left(-R^{KL}\pm\frac{14}{\ell} e^K\wedge e^L\right)=0.
\end{eqnarray}
However, since torsion must be zero by \eqref{tors} this equation does not impose any further restriction on the solution space. This demonstrates the equivalence of our action and Einstein's General Relativity and we conclude that in the case of vacuum we can find actions in which the contact vector $V^A$ is one of the dynamical variables. The extent to which the inclusion of $a_{1}$, $a_{2}$, and $b_{2}$ may complicate the correspondence with vacuum General Relativity is an open question, as is that of the effect of a $V^{2}$ dependence upon coefficients in the action (\ref{actione}) may have. 

\subsubsection{Non-trivial relation between models spacetime and sign of cosmological constant}
Next we consider the equation ${\cal E}_{4I}$ which takes the form
\begin{eqnarray}
0=b_{1}\ell^2\epsilon_{IJKL}\left(e^{J}\wedge R^{KL} \mp\frac{4}{l^{2}} e^{J}\wedge e^{K}\wedge e^{L}\right)  \label{vacein}
\end{eqnarray}
These are the Einstein field equations with cosmological constant. However, while in the case of the MacDowell-Mansouri and Holst actions where the anti-De Sitter/De Sitter model spacetime is associated with a negative/positive cosmological constant (see Appendix \ref{standardform} for details on how the cosmological constant is related to the waywiser radius $\ell$), the relationship in the our case is the {\em opposite} where we have
\begin{eqnarray}
\Lambda_{SO(1,4)} &=& -\frac{12}{l^{2}} \\
\Lambda_{SO(2,3)} &=& +\frac{12}{l^{2}}
\end{eqnarray}
With some hindsight it is perhaps not too surprising that there is no relationship in general between the choice of model spacetime and the sign of the cosmological constant. This should already be clear from the fact that we can add a $c_1$-term, defined above, to the action. 
\section{Conclusions and outlook}\label{conclusions}
In this article we have sought to develop a formulation of Cartan geometry in terms of the notion of idealized waywisers, described on an $n$ dimensional manifold completely in terms of an $SO(p,q)$ connection $A^{AB}$ (where $p+q=n+1$) and a contact point represented by a `contact vector' $V^{A}$. We have called these variables waywiser variables as they encode the response (i.e. the change of contact point) of an idealized waywiser when rolled along paths on the manifold. It was shown that a host of objects familiar from differential geometry, e.g $\Gamma^\rho_{\mu\nu}$, tetrad $e^I_{\mu}$, spin-connection $\omega^{IJ}_{\mu}$, Riemannian curvature $R_{\mu\nu\rho}^{\phantom{\mu\nu\rho}\sigma}$, torsion $T^\rho_{\mu\nu}$, may be recovered from the waywiser variables.

We stressed that General Relativity can be formulated in two distinct ways: one in which the contact vector $V^A$ is treated as a non-dynamical and {\`a priori} postulated object, and a second one in which the contact vector is viewed on a similar footing as the connection $A^{AB}$. To our knowledge the proposed dynamical method in Section \ref{dynapproach} of recovering vacuum General Relativity from an $SO(2,3)$ or $SO(1,4)$ gauge theory is a new one, featuring variations of $V^{E}$ that are unconstrained by Lagrange multipliers. This should be contrasted to previous treatments of actions resembling (\ref{actione}) \cite{Stelle:1979va,Pagels:1983pq}. The non-vanishing value of $V^{2}$ is ensured by the equation \eqref{vev}, rather than more familiar methods such as $V^{2}\neq 0$ corresponding to a local minimum of a potential. This latter possibility was explored in \cite{Wilczek:1998ea} though as part of a framework which breaks diffeomorphism invariance; additionally this approach likely involves a dependence of $V^{2}$ on spacetime coordinates even in the absence of matter fields.

Diffeomorphism invariance is often taken to be the key symmetry group associated with Einstein's General Relativity. However, as noted in Section \ref{nondynapproach}, within the non-dynamical approach diffeomorphism invariance is optional and is broken in generic $SO(p,q)$ gauges. This follows immediately from the fact that $V^A(x)$ is an {\em \`a priori} fixed function on the manifold and therefore explicitly depends on spacetime coordinates $x^\mu$. Only in the particular gauge in which $V^A=\ell\delta^A_4$ is diffeomorphism invariance restored since $V^A$ becomes independent of the spacetime coordinates. In this regard there are two views of the non-dynamical approach that should be considered. One view is that we continue to insist that diffeomorphism invariance should be a fundamental symmetry of nature and in particular of gravitational theories. This would lead to the rejection of the non-dynamical approach in favor of the dynamical one in which diffeomerphism invariance is manifest. Another view would be to reject the idea that diffeomorphism invariace should be regarded as a fundamental symmetry group of gravitational theories. Instead we may adopt the idea that the fundamental symmetry group of gravity is that of `rolling' prescribed by a gauge connection $A^{AB}$ with values in the Lie algebra $\mathfrak{so}(p,q)$. 

Of course, this would require us to understand how matter fields are altered by such a `rolling' and this brings us to the the task of including matter fields within Cartan waywiser geometry. In the spirit of our approach, matter actions must be constructed as integrals of spacetime four-forms constructed from the matter fields and the waywiser variables. Perhaps surprisingly, this appears to be possible at least insofar as recovery of the equations of motion of scalar, spinor, and Yang-Mills fields goes \cite{Pagels:1983pq,Ikeda:2009xb,Westman:2012nn}. 
In this context, the appropriate interpretation of a field $Y^{A}$ is as a \emph{spacetime scalar field} \cite{Pagels:1983pq,Westman:2012nn} (e.g. a Klein-Gordon field). Note that no concept of `inverse-metric' is fundamental at the level of the action here nor does it seem appropriate to require non-degeneracy of the metric since the field equations are valid also in the degenerate cases. Whether these actions may be combined with the action (\ref{actione}) to give a realistic picture of classical gravitation remains an open question.

The geometric interpretation of gravity as Cartan waywiser geometry has hinged on the constancy of the norm $V^{2}$. However, in the presence of general matter content we may imagine that the equations of motion of the variables $V^{A}$ and $A^{AB}$ are sourced such that $V^{2}$ maintains the desired sign but experiences a variation over spacetime: $ V^{2} =  \mp e^{2\phi(x^{\mu})} l^{2}$. Consequently, the metric tensor $g_{\mu\nu}$ takes the following form:
\begin{eqnarray}
g_{\mu\nu} &=& \eta_{AB}D_{\mu}V^{A}D_{\nu}V^{B}\\
            &=& e^{2\phi}\left(\eta_{IJ} e^{I}_{\mu} e^{J}_{\nu} \mp l^{2}\partial_{\mu}\phi \partial_{\nu}\phi\right)
\end{eqnarray}
\\
This amounts to a \emph{disformal} relation between the metric tensor $g_{\mu\nu}$ and the tensor $\eta_{IJ}e^{I}_{\mu}e^{J}_{\nu}$. 
In the present framework, matter is expected to couple to $DV^{A}$ \cite{Pagels:1983pq,Ikeda:2009xb,Westman:2012nn}, and so will couple disformally to the co-tetrad $e^{I}$. The idea of disformal couplings has been an area of recent activity in cosmology 
 \cite{Magueijo:2008sx,Magueijo:2010zc,Zumalacarregui:2010wj,Koivisto:2008ak,Kaloper:2003yf,Bekenstein:2004ne,Skordis:2005xk}; it would be interesting to see whether variation of $V^{2}$ over spacetime may have a phenomenological role.
 
We end by noting that the idea of a waywiser can be generalized to include larger groups. The key feature of a waywiser is that it has a point of contact and a connection that dictates how that point of contact has changed when rolled along some path on the manifold. In this respect it would be interesting to generalize Cartan waywiser geometry to the conformal group $C(1,3)$ which is locally isomorphic to the orthogonal group $SO(2,4)$. 
 
\vspace{0.5cm}
\noindent\textbf{Acknowledgements:} We would like to thank E. Anderson and A. Randono for helpful discussions.
\appendix
\section{Exterior calculus}
Exterior calculus constitutes a powerful tool in differential geometry and this paper makes ample use of it. In order to  make this paper more accessible and self-contained we provide in the following appendices a crash-course in exterior calculus. The various operations, i.e. wedge product, exterior derivative, integration, are defined in such a way that they can be easily understood in terms of tensor operations seen in elementary textbooks in General Relativity.

\subsection{Definition of forms}
In a nutshell, forms are completely anti-symmetric covariant tensors. For example, a scalar $\Phi$ is a zero-form, a connection $A_\mu$ is a one-form, a curvature tensor $F_{\mu\nu}=-F_{\nu\mu}$ is a two-form. In general, we say that a completely antisymmetric covariant tensor of rank $(0,p)$ is a $p$-form. The number $p$ is called the degree of the form. If the manifold dimension is $N$ then no completely antisymmetric covariant tensor exists with more indices than $N$ and consequently no $p$-forms exists if $p>N$. In contradistinction to tensors we see that the number of types of forms is limited by the manifold dimension. Since the index structure of forms is simple and completely specified by its degree $p$ it is convenient to leave out the tensor indices. For example a $p$-form $\Omega_{\mu_1\mu_2\dots\mu_p}$ is written simply as $\Omega$.
\subsection{Exterior algebra}
Next we define a way of multiplying forms together that preserve the antisymmetry. This product is called the {\em wedge product} $\w$. Let $\Omega_1$ and $\Omega_2$ be two forms of degree $p$ and $q$ respectively. Then the wedge product $\Omega_1\w \Omega_2$ is a new form of degree $p+q$. For notational compactness we shall nevertheless omit the symbol $\w$ and simply write $\Omega_1\Omega_2$ since this will not cause any confusion. The basic idea of the wedge product is very simple and can be understood in terms of tensor methods as follows:
\begin{enumerate}
\item Write the forms as covariant tensors: $\Omega_{1\mu_1\mu_2\dots\mu_p}$ and $\Omega_{2\mu_1\mu_2\dots\mu_q}$
\item Multiply them as tensors: $\Omega_{1\nu_1\dots\nu_p}\Omega_{2\nu_{p+1}\dots\nu_{p+q}}$
\item Antisymmetrize: $\frac{(p+q)!}{p!q!}\Omega_{[1\nu_1\dots\nu_p}\Omega_{2\nu_{p+1}\dots\nu_{p+q}]}$.
\end{enumerate}
The last object defines the $p+q$-form $\Omega_1\Omega_2$ with tensor indices explicit. The following formal properties of the wedge product can easily be deduced. Let $\Omega_1$, $\Omega_2$, and $\Omega_3$ be a $p$-form, $q$-form, and $r$-form respectively, and $\alpha$ and $\beta$ real- or complex numbers.
\begin{itemize}
\item Linearity: $(\alpha\Omega_1+\beta\Omega_2) \Omega_3=\alpha\Omega_1\Omega_3+\beta\Omega_2\Omega_3$
\item Commutation law: $\Omega_1\Omega_2=(-1)^{pq}\Omega_2\Omega_1$ where $\Omega_1$ is a $p$-form and $\Omega_2$ is a $q$-form.
\item Associativity: $\Omega_1(\Omega_2\Omega_3)=(\Omega_1\Omega_2)\Omega_3$
\end{itemize}
The wedge-product of the two forms $\Omega_1$ and $\Omega_2$, of degree $p$ and $q$ say, produces a new form $\Omega_3=\Omega_1\Omega_2$ of degree $p+q$. Thus, if $p+q>N$ then $\Omega_1\Omega_2\equiv0$. The above rules defines the exterior algebra of forms.
\subsection{Coordinate basis}
A coordinate system is a collection of $N$ scalar fields $x^\mu=(x_1,\dots,x_N)$ on an $N$-dimensional manifold $\m M$. The gradients of these scalars $dx^\mu$ forms a set of $N$ one-forms which are normals to the equipotential surfaces $x^\mu=const$ for $\mu=1,2,\dots,N$. These normals are nothing but the gradients of the coordinate zero-forms $x^1,x^2,\dots$. As such they have one lowercase index and are therefore examples of one-forms. We write them $dx^1,dx^2,\dots,dx^N$ where the $d$ is here understood as a gradient. As such $dx^\mu$ are not infinitesimals. 

These one-forms collectively written as $dx^\mu$ are a set of co-vectors that span space of one-forms. Thus we can expand a one-form in terms of its coordinate coefficients $A_\mu$ as $A=A_\mu dx^\mu$. Similarly, the objects $dx^\mu\wedge dx^\nu\equiv dx^\mu dx^\nu$ are two-forms and they span the space of two forms. A two-form can then be expanded in terms of its coordinate coefficients $F_{\mu\nu}$ as $F=\frac{1}{2}F_{\mu\nu}dx^\mu dx^\nu$. More generally, any $p$-form $\Omega$ can be expanded in the coordinate one-form basis as follows
\begin{eqnarray}
\Omega=\frac{1}{p!}\Omega_{\mu_1\dots\mu_p}dx^{\mu_1} dx^{\mu_2}\dots dx^{\mu_p}.
\end{eqnarray}
Instead of forming the gradient of each scalar $x^\mu$ we can also consider the tangent vectors to the coordinate defined by varying one coordinate while holding all the others fixed. This yields $N$ tangent vectors which we here shall denote $\partial_\mu$ which then forms a set of basis vectors on the tangent space. Thus a vector may be written as
\begin{align}
V=V^\mu\partial_\mu.  
\end{align}
A general $(p,q)$ tensor $T$ is then expanded in the coordinate basis as 
\begin{align}
T=T^{\mu_1\dots\mu_p}_{\nu_1\dots\nu_q}dx^{\nu_1}\otimes\dots \otimes dx^{\nu_q}\otimes \partial_{\mu_1}\otimes \dots \otimes \partial_{\mu_p}
\end{align}
That we use the symbol $\partial_\mu$ which also denotes a partial derivative is no accident. The partial derivative is defined to take the derivative along the direction defined by changing the specific coordinate $x^\mu$ while holding the values $x^\nu$, $\nu\neq\mu$, of all other fixed. Thus, a vector has then a natural action on a scalar field $\phi$ by 
\begin{equation}
V(\phi)=V^\mu\partial_{\mu}\phi
\end{equation}
Note, however, that a general contravariant tensor $T=T^{\mu_1\dots\mu_p}\partial_{\mu_1}\otimes\dots\otimes \partial_{\mu_p}$ does not have a natural coordinate independent action on a scalar. 

\subsection{Duality between forms and antisymmetric contravariant tensor densities}\label{duality} 
There is however another form of duality which always exists: The completely antisymmetric Levi-Civita tensor density $\varepsilon^{\mu_1\mu_2\dots\mu_N}$ establishes an isometry between the space of $p$-forms and the space of completely antisymmetric $(N-p,0)$-rank tensor densities of weight $+1$. We will use the symbol $\sim$ to denote the dual quantity. Specifically, let $\Omega$ be some $p$-form, then the dual contravariant antisymmetric $+1$ tensor density $\Omega^{\mu_{p+1}\dots\mu_N}$ is defined as

\begin{eqnarray}
\Omega=\frac{1}{p!}\Omega_{\mu_1\dots\mu_p}dx^{\mu_1}\dots dx^{\mu_p}\sim\frac{1}{p!}\Omega_{\mu_1\dots\mu_p}\varepsilon^{\mu_1\dots\mu_p\dots\mu_N}
\end{eqnarray}
where the two notationally distinct Levi-Civita symbols $\epsilon$ and $\varepsilon$ are defined so that 
\begin{eqnarray}
\epsilon_{\mu_1\mu_2\dots \mu_N}\varepsilon^{\mu_1\mu_2\dots \mu_N}=+N!.
\end{eqnarray}
As a simple concrete example we can see that, in the case of four spacetime dimensions, the object dual to the four-form $\m E=\frac{1}{4!}\epsilon_{IJKL}e^I\w e^J\w e^K\w e^L$, is nothing but the usual scalar density volume element $e\equiv det(e^I_\mu)$, i.e. we have
\begin{eqnarray}
\m E=\frac{1}{4!}\epsilon_{IJKL}e^Ie^Je^Ke^L&=&\frac{1}{4!}\epsilon_{IJKL}e^I_\mu e^J_\nu e^K_\rho e^L_\sigma dx^\mu dx^\nu dx^\rho dx^\sigma\nonumber\\
&\sim& \varepsilon^{\mu\nu\rho\sigma}\frac{1}{4!}\epsilon_{IJKL}e^I_\mu e^J_\nu e^K_\rho e^L_\sigma\equiv det(e^I_\mu)=e
\end{eqnarray}
This duality between differential forms and contravariant antisymmetric tensor densities is useful since it allows us to translate between expressions written in differential forms forms and the more common tensorial notation which is more common within the physics community.
\subsection{Forms as linear functionals}
A $p$-form written as $\Omega=\frac{1}{p!}\Omega_{\mu_1\dots\mu_p}dx^{\mu_1}\dots dx^{\mu_p}$ should not be interpreted as an infinitesimal quantity despite the appearance of the $dx^\mu$'s which might naively be interpreted as infinitesimal displacements which we in this paper instead denote as $\delta x^\mu$. Rather, forms are to be understood as completely antisymmetric multi-linear functionals $\Omega:T_p({\cal M})\otimes\dots\otimes T_p({\cal M})\rightarrow \mb R$. For example, a one-form $A$ fed a vector $V$ yields the real number $A(V)$. If $A$ happens to be the exterior derivative of a scalar then we may note the following identity
\begin{align}
d\phi(V)=\partial_\mu\phi V^\mu=V^\mu\partial_\mu\phi=V(\phi)
\end{align}
We see that the coordinate basis one-form $dx^\mu$ fed the coordinate basis vector $\partial_\nu$ yields 
\begin{align}
dx^\mu(\partial_\nu)=\partial_\nu (x^\mu)=\frac{\partial x^\mu}{\partial x^\nu}=\delta^\mu_\nu
\end{align}
Thus we have $A(V)=A_\mu (dx^\mu)(V)=A_\mu V^\nu(dx^\mu)(\partial_\nu)=A_\mu V^\nu\delta^\mu_\nu=A_\mu V^\mu$.

The coordinate basis p-form $dx^{\mu_1}\dots dx^{\mu_p}$ fed $p$ coordinate basis vectors yields
\begin{align}
dx^{\mu_1}\dots dx^{\mu_p}(\partial_{\nu_1},\dots,\partial_{\nu_p})=\delta^{\mu_1\dots\mu_p}_{\nu_1\dots\nu_p}=\frac{1}{(N-p)!}\epsilon_{\nu_1\dots\nu_p\rho_{p+1}\dots \rho_N}\varepsilon^{\mu_1\dots\mu_p\rho_{p+1}\dots \rho_N}
\end{align}
so that we have
\begin{align}
\Omega(V_1,\dots,V_p)&=\frac{1}{p!}\Omega_{\mu_1\dots\mu_p}(dx^{\mu_1}\dots dx^{\mu_p})(V_1,\dots, V_p)=\frac{1}{p!}\Omega_{\mu_1\dots\mu_p}V_1^{\nu_1}\dots V_p^{\nu_p}(dx^{\mu_1}\dots dx^{\mu_p})(\partial_{\nu_1},\dots,\partial_{\nu_p})\nn\\
&=\frac{1}{p!(N-p)!}\epsilon_{\nu_1\dots\nu_p\rho_{p+1}\dots \rho_N}\varepsilon^{\mu_1\dots\mu_p\rho_{p+1}\dots \rho_N}\Omega_{\mu_1\dots\mu_p}V_1^{\nu_1}\dots V_p^{\nu_p}=\frac{1}{p!}\delta^{\mu_1\dots\mu_p}_{\nu_1\dots\nu_p}\Omega_{\mu_1\dots\mu_p}V_1^{\nu_1}\dots V_p^{\nu_p}\nn
\end{align}
The collection of vectors $(V_1,\dots, V_p)$ forms a $p$-dimensional parallelepiped in an $N$-dimensional tangent space.
\subsection{Exterior differentiation}

Next we define a coordinate independent derivative operator, called the {\em exterior derivative}, for forms that preserve the complete antisymmetry and generates from a $p$-form $\Omega$ a new form $d\Omega$ with degree $p+1$. The partial derivative $\partial_\mu$ will not do since: 1) it is coordinate dependent when acting on a $p$-form with $p>0$ and 2) it takes us out of the space of forms, i.e. completely antisymmetric tensors. The basic idea of the exterior derivative is simple and amounts to carrying out the following steps.
\begin{enumerate}
\item Write the form as a covariant tensor: $\Omega_{\mu_1\mu_2\dots\mu_p}$
\item Take the partial derivative: $\partial_{\mu_{p+1}}\Omega_{\mu_1\mu_2\dots\mu_p}$
\item Antisymmetrize: $(p+1)\partial_{[\mu_{p+1}}\Omega_{\mu_1\mu_2\dots\mu_p]}$.
\end{enumerate}
The last completely antisymmetric covariant vector defines the exterior derivative denoted $d\Omega$. This object is coordinate independent. The following formal properties of the exterior derivative can easily be checked:
\begin{itemize}
\item The components of the exterior derivative of a zero-form are its partial derivatives $(d\Phi)_\mu=\partial_\mu \Phi$ ($d\Phi=\partial_{\mu}\Phi dx^{\mu}$).
\item Linearity: $d(\alpha\Omega_1+\beta\Omega_2)=\alpha d\Omega_1+\beta d\Omega_2$
\item Leibniz rule: $d(\Omega_1\Omega_2)=d\Omega_1\Omega_2+(-1)^{p}\Omega_1 d\Omega_2$ where $\Omega_1$ is a $p$-form.
\item $d^2\Omega=d(d\Omega)\equiv0$ for all $p$-forms $\Omega$ and all $p$.
\end{itemize}
The factor of $(-1)^p$ in the Leibniz rule is there to compensate for the commutation rule for forms. The last property is nothing but a restatement of the commutativity of partial derivatives. The exterior derivative of an $N$-form is automatically zero since there are no forms with degree $N+1$.

\subsection{Integration of forms}

Consider an integral of some quantity on some $p$-dimensional surface in an $N$-dimensional space. If the surface is parametrized by $\xi^\alpha$ so that $x^\mu=x^\mu(\xi)$ consists of points in that surface such integral is written as
\begin{align}
\int \Phi(x(\xi))\delta^p\xi.
\end{align}
We write $\delta^{p}\xi$ instead of the standard $d^{p}\xi$ so as to avoid confusion with the `$d$' appearing in the formalism of differential forms. The appearance of $\delta^p\xi$ and $x(\xi)$ clearly shows that this integral is not written in a manifestly coordinate and parametrization independent manner. The language of differential forms allows for a neat coordinate and parametrization-free notation. In fact, forms are precisely those elementary mathematical objects which appear under integral signs. A one-form $A$ can be integrated along a one-dimensional curve on the manifold, a two form $F$ over a two-dimensional surface, and a $p$-form $\Omega$ over a $p$-dimensional sub-manifold $\Sigma$.

To see this clearly and establish a concrete connection with standard notation let us consider the integral of some $p$-form $\Omega$ on some $p$-dimensional surface. Although we write $\int \Omega$ one should not fall prey to the temptation of thinking of $\Omega$ as an infinitesimal quantity. Rather we should think of evaluating the integration of the $p$-form $\Omega$ on a $p$-dimensional surface $\Sigma$ in the following way. Using $x^\mu$ as coordinates on $\m M$ and some parametrization $\xi^i$ for the surface $\Sigma$ we can span the tangent space $T_\xi(\Sigma)$ by the vectors
\begin{align}
\overrightarrow{\xi^1}=\frac{\partial x^\mu}{\partial \xi^1}\partial_\mu\qquad \overrightarrow{\xi^2}=\frac{\partial x^\mu}{\partial \xi^2}\partial_\mu\qquad \dots\qquad \overrightarrow{\xi^p}=\frac{\partial x^\mu}{\partial \xi^p}\partial_\mu
\end{align}
From these we can now define $p$ infinitesimal displacement vectors 
\begin{align}
\overrightarrow{\delta\xi^1}=\delta\xi^1\frac{\partial x^\mu}{\partial \xi^1}\partial_\mu\qquad \overrightarrow{\delta\xi^2}=\delta\xi^2\frac{\partial x^\mu}{\partial \xi^2}\partial_\mu\qquad \dots\qquad \overrightarrow{\delta\xi^p}=\delta\xi^p\frac{\partial x^\mu}{\partial \xi^p}\partial_\mu
\end{align}
where $\delta \xi^i$, $i=1,\dots,p$ are infinitesimals. This collection of infinitesimal vectors forms a $p$-dimensional parallelepiped. We can now form at each point on $\Sigma$ an infinitesimal real number by feeding the form $\Omega$ the infinitesimal parallelepiped $(\overrightarrow{\delta\xi^1},\dots,\overrightarrow{\delta\xi^p})$, i.e. $\Omega(\overrightarrow{\delta\xi^1},\dots,\overrightarrow{\delta\xi^p})$. The evaluation of the integral $\int \Omega$ then simply consists of summing all these infinitesimal real numbers together. Specifically, the evaluation goes as follows:
\begin{align}
\int \Omega&=\int \Omega(\overrightarrow{\delta\xi^1},\dots,\overrightarrow{\delta\xi^p})=\int\frac{1}{p!}\Omega_{\mu_1\dots\mu_p}(dx^{\mu_1}\dots dx^{\mu_p})(\overrightarrow{\delta\xi^1},\dots,\overrightarrow{\delta\xi^p})\nn\\
&=\int \frac{1}{p!}\delta\xi^1\frac{\partial x^{\nu_1}}{\partial \xi^1}\dots \delta\xi^p\frac{\partial x^{\nu_p}}{\partial \xi^p}\Omega_{\mu_1\dots\mu_p}(dx^{\mu_1}\dots dx^{\mu_p})(\partial_{\nu_1},\dots,\partial_{\nu_p})\nn\\
&=\int \frac{1}{p!}\frac{\partial x^{\nu_1}}{\partial \xi^1}\dots\frac{\partial x^{\nu_p}}{\partial \xi^p}\Omega_{\mu_1\dots\mu_p}\delta^{\mu_1\dots\mu_p}_{\nu_1\dots\nu_p}\delta^p\xi\nn\\
&=\int \frac{1}{p!(N-p)!}\frac{\partial x^{\nu_1}}{\partial \xi^1}\dots\frac{\partial x^{\nu_p}}{\partial \xi^p}\Omega_{\mu_1\dots\mu_p}\epsilon_{\nu_1\dots\nu_p\rho_{p+1}\dots\rho_N}\varepsilon^{\mu_1\dots\mu_p\rho_{p+1}\dots\rho_N}\delta^p\xi.
\end{align}
We note again that the coordinate volume element $\delta^p\xi$ is usually written as $d^p\xi$ but here we have used the symbol $\delta$ rather than $d$ so as to not confuse it with the exterior derivative symbol which appears in $dx^\mu$ for example. 

For concreteness let us consider a standard flux integral over a two-dimensional surface in a three-dimensional flat Euclidean space. A typical notation for this is
\begin{align}
\Phi=\int B\cdot n\delta A
\end{align}
where $B$ is some vector field, $n$ the field of normals on the surface, and $\delta A$ the area element. We write $\delta A$ rather than the standard $dA$ to avoid confusion with the exterior derivative symbol $d$. To compute the normal $n$ and area element $\delta A$ we first parametrize the surface $X(u,v)=(x(u,v),y(u,v),z(u,v))$ and then compute the tangent vectors 
\begin{align}
X_u=\frac{\partial X^{i}}{\partial u}\partial_{i}\qquad X_v=\frac{\partial X^{i}}{\partial v}\partial_{i}
\end{align}
and defining the infinitesimal vectors
\begin{align}
\overrightarrow{\delta u}=\delta u\frac{\partial X^i}{\partial u}\partial_i\qquad \overrightarrow{\delta v}=\delta v\frac{\partial X^i}{\partial v}\partial_i
\end{align}
so that the normal and area element become
\begin{align}
n=\frac{X_u\times X_v}{|X_u\times X_v|}\qquad \delta A=|X_u\times X_v|\delta u\delta v
\end{align}
where it may be checked explicitly that indeed $\delta A$ is the area of a parallelepiped spanned by vectors $\overrightarrow{\delta u}$ and $\overrightarrow{\delta u}$. The flux integral now reads
\begin{align}
\Phi=\int B\cdot (X_u\times X_v)\delta u\delta v \label{flux1}
\end{align}
where the dot denotes the metric inner-product $\delta_{ij}B^{i}(X_u\times X_v)^{j}$. Indeed, in components (\ref{flux1}) reads
\begin{align}
\int B^i(X_u\times X_v)_i\delta u\delta v=\int \frac{1}{2}\varepsilon^{ijk}F_{kl}\epsilon_{imn}\frac{\partial X^m}{\partial u}\frac{\partial X^n}{\partial v}\delta u\delta v
\end{align}
where we have introduce the dual antisymmetric object $F_{ij}$ defined via $B^i=\frac{1}{2}\varepsilon^{ijk}F_{jk}$. Using the identity
\begin{align}
(dx^jdx^k)(\partial_m,\partial_n)=\delta^{jk}_{mn}=\delta^j_m\delta^k_n-\delta^j_n\delta^k_m=\varepsilon^{ijk}\epsilon_{imn}
\end{align}
we get
\begin{align}
\Phi=\int \frac{1}{2}F_{kl}(dx^jdx^k)(\partial_m,\partial_n)\frac{\partial X^m}{\partial u}\frac{\partial X^n}{\partial v}\delta u\delta v=\int F(\partial_m,\partial_n)\frac{\partial X^m}{\partial u}\frac{\partial X^n}{\partial v}\delta u\delta v
\end{align}
hence we have 
\begin{align}
\Phi=\int F(\overrightarrow {\delta u},\overrightarrow {\delta v})=\int F
\end{align}
We note that the orientation is specified by the normal $n$ and that this orientation is automatically accounted for in the forms language. We also note that a flux integral in coordinate and parameterization independent language naturally involves the two-form $F=\frac{1}{2}F_{ij}dx^idx^j$ rather then the vector $F^i$. It is in this precise sense we may say that forms `{\em are the things which occur under integral signs.}' \cite{flanders2012differential}. 
\section{Gauge connections, curvature, and Bianchi identities}
We provide here a brief exposition of the basic techniques and ideas of gauge connections in the language of forms. Although the formulas of this section is valid for any gauge group we will mostly use the waywiser variables to illustrate the ideas. 

The contact vector $V^A$ appears with a gauge index $A$ and transforms under a spacetime-dependent gauge transformation as $V^A\rightarrow \theta(x)^A_{\ph AB}V^B$. Objects with gauge index downstairs, e.g. $U_A$, transforms as $U_A\rightarrow U_B(\theta^{-1})^B_{\ph BA}$ so that $U_AV^A$ is invariant under arbitrary gauge transformations. This fixes the transformation law of mixed objects $W^A_{\ph AB}$ as $W^A_{\ph AB}\rightarrow \theta^A_{\ph AC}W^C_{\ph CD}(\theta^{-1})^D_{\ph DB}$.

The exterior derivative of $dV^A$ transforms inhomogeneously $d(\theta^A_{\ph AB}V^B)\neq \theta^A_{\ph AB}dV^B$ and $dV^A$ under a spacetime-dependent gauge transformation $V^A\rightarrow \theta(x)^A_{\ph AB}V^B$. It is therefore not a gauge-covariant object. In order to restore gauge-covariance the exterior derivative is replaced by the gauge covariant exterior derivative $d\rightarrow D^{(A)}$:
\begin{eqnarray}
D^{(A)}V^A\equiv dV^A+A^A_{\ph AB}V^B \qquad D^{(A)}U_A\equiv dU_A-A^B_{\ph BA}U_B 
\end{eqnarray} 
with the minus sign on the right equation guaranteeing that $D(U_AV^A)=d(U_AV^A)$. The requirement of gauge-covariance, i.e. $D^{(A')}(\theta^A_{\ph AB}V^B)=\theta^A_{\ph AB}D^{(A)}V^B$, implies immediately that the connection $A^A_{\ph AB}$ transforms inhomogeneously under local gauge transformation:
\begin{eqnarray}
A^A_{\ph AB}\rightarrow A^{\prime A}_{\ph{\prime A}B}=-d\theta^A_{\ph AC}(\theta^{-1})^C_{\ph CB}+\theta^A_{\ph AC}A^C_{\ph CD}(\theta^{-1})^D_{\ph DB}.
\end{eqnarray}
We will often write $D$ for the gauge-covariant instead of the more cumbersome notation $D^{(A)}$ wherever no confusion can arise. The gauge covariant exterior derivative of some $p$-form, $\Omega^A_{\ph AB}$ say, is given by
\begin{eqnarray}
D\Omega^A_{\ph AB}=d\Omega^A_{\ph AB}+A^A_{\ph AC}\wedge \Omega^C_{\ph CB}-A^C_{\ph CB}\wedge \Omega^A_{\ph AC}
\end{eqnarray}
The curvature two-form $F^A_{\ph AB}$ defined by
\begin{eqnarray}\label{standef}
F^A_{\ph AB}\equiv dA^A_{\ph AB}+A^A_{\ph CC}\wedge A^C_{\ph CB}
\end{eqnarray}
can straightforwardly be shown to transform as $F^A_{\ph AB}\rightarrow \theta^A_{\ph AC}F^C_{\ph CD}(\theta^{-1})^D_{\ph DB}$ and is therefore gauge covariant. Note however that the gauge covariant derivative applied to the (gauge non-covariant) connection 
\begin{eqnarray}
DA^{AB}=dA^{AB}+A^A_{\ph{A}C}\wedge A^{CB}+A^B_{\ph{B}C}\wedge A^{AC}=F^{AB}+A^A_{\ph{A}C}\wedge A^{CB}
\end{eqnarray}
is not gauge covariant. 

The identity $DF^A_{\ph AB}\equiv 0$ is extremely useful and is called the {\em first Bianchi identity}. It follows immediately from the definition of the gauge-covariant exterior derivative and the rules of exterior calculus:
\begin{multline}
DF^A_{\ph AB}\equiv D^2A^A_{\ph AB}\equiv dF^A_{\ph AB}+A^A_{\ph AC}\wedge F^C_{\ph CB}-A^C_{\ph CB}\wedge F^A_{\ph AC}=d(dA^A_{\ph AB}+A^A_{\ph AC}\wedge A^C_{\ph CB})\nonumber\\
+A^A_{\ph AC}\wedge (dA^C_{\ph CB}+A^C_{\ph CD}\wedge A^D_{\ph DB})-A^C_{\ph CB}\wedge (dA^A_{\ph AC}+A^A_{\ph AD}\wedge A^D_{\ph DC})\\
=dA^A_{\ph AC}\wedge A^C_{\ph CB}-A^A_{\ph AC}\wedge dA^C_{\ph CB}+A^A_{\ph AC}\wedge dA^C_{\ph CB}
+A^A_{\ph AC}\wedge A^C_{\ph CD}\wedge A^D_{\ph DB}-A^C_{\ph CB}\wedge dA^A_{\ph AC}-A^C_{\ph CB}\wedge A^A_{\ph AD}\wedge A^D_{\ph DC}\nonumber\\
=dA^A_{\ph AC}\wedge A^C_{\ph CB}+A^A_{\ph AC}\wedge A^C_{\ph CD}\wedge A^D_{\ph DB}-dA^A_{\ph AC}\wedge A^C_{\ph CB} -A^A_{\ph AD}\wedge A^D_{\ph DC}\wedge A^C_{\ph CB}\equiv0\nonumber
\end{multline}
By taking the gauge-covariant derivative of the torsion tensor defined by $T^A\equiv F^A_{\ph AB}V^B$ and making use of the Leibniz rule and the first Bianchi identity $DF^A_{\ph AB}\equiv0$ we obtain the {\em second Bianchi identity}
\begin{eqnarray}
DT^A\equiv D(F^A_{\ph AB}V^B)=F^A_{\ph AB}\wedge DV^B
\end{eqnarray}
\section{The Palatini action in the language of forms}
\label{tensorformtrans}
\label{palappendix}
To help make contact with standard notation we illustrate how the Palatini action of the Einstein-Cartan theory (written as a four-form) corresponds to the more familiar Einstein-Hilbert action (written in terms of the density $\sqrt{-g}$ and coordinate displacement product $d^{4}x\equiv \delta^{4}x$). The Palatini action is as follows: 
\begin{eqnarray}
{\cal S}_P=\int \epsilon_{IJKL}e^I e^J R^{KL}.
\end{eqnarray}
The one-form $e^I$ is the co-tetrad (the inverse tetrad) and $R^{IJ}$ is the Riemann curvature two-form defined by  $R^{IJ}=d\omega^{IJ}+\omega^I_{\ph IK} \omega^{KJ}$ with $\omega^{IJ}$ a  one-form valued in the Lie algebra of $SO(1,3)$. 

This action is written in a manifestly coordinate independent way. In order to relate this action to the more well-known Einstein-Hilbert action $\int \sqrt{-g}Rd^4x$ which is not written in a manifestly coordinate independent way we must introduce a coordinate system, $x^\mu$ say. We can now expand the forms $e^I$ and $R^{IJ}$ in the basis $dx^\mu$: $e^I=e^I_\mu dx^\mu$ and $R^{KL}=\frac{1}{2}R_{\mu\nu}^{\ph {\mu\nu}KL}dx^\mu dx^\nu$. Thus we have,
\begin{eqnarray}
{\cal S}_P&=&\int \epsilon_{IJKL}e^I e^J R^{KL}=\int \frac{1}{2}\epsilon_{IJKL}e_\mu^I e_\nu^J R_{\rho\sigma}^{\ph{\rho\sigma}KL}dx^\mu dx^\nu dx^\rho dx^\sigma\nn
\end{eqnarray}
Next-in terms of displacements in local coordinates $(x^{0},x^{1},x^{2},x^{3})$- we define the infinitesimal four-dimensional parallelepiped\footnote{Since we are integrating over all of the four-dimensional manifold $\m M$ rather than some subsurface we have without loss of generality let the parametrization $\xi$ coincide with the coordinates $x$.}
\begin{align}
\overrightarrow{\delta x^0}=\delta x^0\frac{\partial x^\nu}{\partial x^0}\partial_\nu=\delta x^0\partial_0;\quad \overrightarrow{\delta x^1}=\delta x^1\frac{\partial x^\nu}{\partial x^1}\partial_\nu=\delta x^1\partial_1;\quad\overrightarrow{\delta x^2}=\delta x^2\frac{\partial x^\nu}{\partial x^2}=\delta x^2\partial_2;\quad\overrightarrow{\delta x^3}=\delta x^3\frac{\partial x^\nu}{\partial x^3}\partial_\nu=\delta x^3\partial_3
\end{align}
, which when fed to the four-form $\epsilon_{IJKL}e^Ie^JR^{KL}$ yields
\begin{align}
(\epsilon_{IJKL}e^Ie^JR^{KL})(\overrightarrow{\delta x^0},\overrightarrow{\delta x^1},\overrightarrow{\delta x^2},\overrightarrow{\delta x^3})&=\frac{1}{2}\epsilon_{IJKL}e_\mu^I e_\nu^J R_{\rho\sigma}^{\ph{\rho\sigma}KL}(dx^\mu dx^\nu dx^\rho dx^\sigma)(\overrightarrow{\delta x^0},\overrightarrow{\delta x^1},\overrightarrow{\delta x^2},\overrightarrow{\delta x^3})\nn\\
&=\frac{1}{2}\epsilon_{IJKL}e_\mu^I e_\nu^J R_{\rho\sigma}^{\ph{\rho\sigma}KL}\delta^{\mu\nu\rho\sigma}_{0123}\delta^4x=\int \frac{1}{2}\epsilon_{IJKL}e_\mu^I e_\nu^J R_{\rho\sigma}^{\ph{\rho\sigma}KL}\varepsilon^{\mu\nu\rho\sigma}\underbrace{\epsilon_{0123}}_{=+1}\delta^4x\nn\\
&=\frac{1}{2}\epsilon_{IJKL}e_\mu^I e_\nu^J R_{\rho\sigma}^{\ph{\rho\sigma}KL}\varepsilon^{\mu\nu\rho\sigma}\delta^4x\nn
\end{align}
In order to see that the action ${\cal S}_P$ is nothing but (twice) the Einstein-Hilbert action $\m S_{EH}$ written in the variables $e^I$ and $\omega^{IJ}$ we do the following rewriting
\begin{eqnarray}
\m S_{P}&=&\int\epsilon_{IJKL}e^Ie^JR^{KL}=\int \frac{1}{2}\epsilon_{IJMN}e_\mu^I e_\nu^J e^M_\kappa e^N_\tau e^\kappa_Ke^\tau_L R_{\rho\sigma}^{\ph{\rho\sigma}KL}\varepsilon^{\mu\nu\rho\sigma}\delta^4x\nonumber\\
&=&\int\frac{1}{2}\epsilon_{IJKL}e_\mu^I e_\nu^J R_{\rho\sigma}^{\ph{\rho\sigma}KL}\varepsilon^{\mu\nu\rho\sigma}\delta^4x=\int \frac{1}{2}e\epsilon_{\mu\nu\kappa\tau}\varepsilon^{\mu\nu\rho\sigma}e^\kappa_Ke^\tau_L R_{\rho\sigma}^{\ph{\rho\sigma}KL}\delta^4x\nonumber\\
&=&\int \frac{1}{2}e2(\delta^\rho_\tau\delta^\sigma_\kappa-\delta^\rho_\kappa\delta^\sigma_\tau)e^\kappa_Ke^\tau_L R_{\rho\sigma}^{\ph{\rho\sigma}KL}\delta^4x=\int 2e e^\mu_Ie^\nu_J R_{\mu\nu}^{\ph{\mu\nu}IJ}\delta^4x\nn\\
&=&\int 2\sqrt{-g} R\delta^4x=2\m S_{EH}\nonumber
\end{eqnarray}
where we made use of the identities
\begin{eqnarray}
\sqrt{-g}=e\quad R=e^\mu_Ie^\nu_J R_{\mu\nu}^{\ph{\mu\nu}IJ}\quad\epsilon_{\mu\nu\kappa\tau}\varepsilon^{\mu\nu\rho\sigma}=
2(\delta^\rho_\kappa\delta^\sigma_\tau-\delta^\rho_\tau\delta^\sigma_\kappa)\quad e^\mu_I e_\mu^J=\delta^J_I\quad e\epsilon_{\mu\nu\rho\sigma}=\epsilon_{IJKL}e^I_\mu e^J_\nu e^K_\rho e^L_\sigma\nonumber
\end{eqnarray}
with $e$ the co-tetrad determinant and $e^\mu_I$ its inverse. As before we have written $\delta^4x$ rather than $d^4x$ as to not confuse it with the symbol $d$ for the exterior derivative. 
\section{The variational calculus of differential forms}\label{variationalforms}
A spacetime action ${\cal S}$ is per definition an integral ${\cal S}=\int\mathcal{L}$ of some four-form ${\cal L}$ over some spacetime region $V$. Since all the basic variables in Cartan waywiser geometry are themselves differential forms, and the equations of motions are obtained by requiring the action to be extremized, we provide, for completeness and accessibility, an exposition of the variational calculus of differential forms and related helpful tricks which simplify calculations immensely. For the sake of simplicity, our Lagrangian four-forms $\mathcal{L}$ will be assumed to be polynomial in the basic forms.

The variation of a p-form $\Omega$ is as usual defined as $\Omega\rightarrow \Omega+\delta\Omega$. The variation symbol $\delta$ commutes with the exterior derivative $\delta d\Omega=d\delta\Omega$ which follows immediately from the linear property of the exterior derivative: $\delta d\Omega\equiv d(\Omega+\delta\Omega)-d\Omega =d\Omega+d\delta\Omega-d\Omega=d\delta\Omega$.

Let us now consider some action ${\cal S}=\int_V \mathcal{L}$ where $\mathcal{L}$ is a four-form that for concreteness depends on some form $\Omega$ and it's first exterior derivative $d\Omega$, i.e. $\mathcal{L}=\mathcal{L}(\Omega,d\Omega)$. In order to obtain the equations of motion for $\Omega$ we wish to vary the action with respect to the differential form $\Omega$. The variation $\delta_\Omega {\cal S}$ is defined by
\begin{eqnarray}
\delta_\Omega {\cal S}=\int_V \delta_\Omega \mathcal{L}(\Omega,d\Omega)\equiv \int_V \mathcal{L}(\Omega+\delta\Omega,d\Omega+d\delta\Omega)-\mathcal{L}(\Omega,d\Omega)=\int_V\mathcal{L}(\delta\Omega,d\Omega)+\mathcal{L}(\Omega,d\delta\Omega)
\end{eqnarray}
In order to extract equations of motion we as usual integrate by parts which we now turn to. 
\subsection{Integration by parts}
After a variation of a Lagrangian four-form $\mathcal{L}$ with respect to a form $\Omega$ we might end up with terms like $d(\delta_\Omega\omega)$ where $\omega$ is some three-form. If we now assume that the variation of $\Omega$ is zero at the boundary $\partial V$, i.e. $\delta\Omega|_{\partial V}=0$, we also have that $\delta_\Omega\omega|_{\partial V}=0$. Gauss theorem then yields
\begin{eqnarray}
\int_V \delta_\Omega d(\omega)=\int_V d(\delta_\Omega\omega)=\int_{\partial V} \delta_\Omega\omega=0
\end{eqnarray}
and we conclude that terms like in a Lagrangian which are a exterior derivatives of a three-forms, e.g. $d\omega$ above, do not alter the equations of motion. These are also called topological terms. 

Suppose now that we have obtained
\begin{eqnarray}
\int_V \delta\Omega \Psi + d\delta\Omega\Phi
\end{eqnarray}
after a variation with respect to $\Omega$. By making use of the Leibniz rule for exterior derivatives 
\begin{eqnarray}
d(\delta\Omega\Phi)=d\delta\Omega\Phi+(-1)^p\delta\Omega d\Phi
\end{eqnarray}
we see that we can simplify the above variation using Gauss theorem and the fact that the variation $\delta\Omega$ vanishes at the boundary
\begin{align}
\int_V \delta\Omega \Psi + d\delta\Omega\Phi&=\int_V \delta\Omega \Psi + d(\delta\Omega\Phi)-(-1)^p\delta\Omega d\Phi\nonumber\\
&=\int_V \delta\Omega \Psi-(-1)^p\delta\Omega d\Phi + \underbrace{\int_{\partial V} \delta\Omega\Phi}_{=0}\nonumber\\&=\int_V \delta\Omega(\Psi-(-1)^p d\Phi)\nonumber
\end{align}
If the action is supposed to extremized its variation must be zero for all choices of $\delta\Omega$. This means that 
\begin{eqnarray}
\Psi-(-1)^p d\Phi=0
\end{eqnarray}
which then constitute the equations of motion.
\subsection{Methods using the gauge covariant exterior derivative}
We can now extend the above discussion to include gauge covariant exterior derivatives $D$. Strictly speaking there is no need to do this but it simplifies calculations immensely and keeps the expressions manifestly gauge covariant throughout the calculation. 

For concreteness we use the waywiser forms and their gauge-covariant derivatives to illustrate the computational techniques involved. As in the case of the exterior derivative, we infer from linearity that the variation symbol $\delta$ commutes with the gauge covariant exterior derivative $D$. In the case of the curvature two-form we have the important relation
\begin{eqnarray}
\delta_A F^{AB}=\delta_A (dA^{AB}+A^A_{\ph AC}\wedge A^{CB})=d\delta A^{AB}+\delta A^A_{\ph AC}\wedge A^{CB}+A^A_{\ph AC}\wedge \delta A^{CB}=D\delta A^{AB}
\end{eqnarray}
Because the gauge covariant exterior derivative satisfies the Leibnitz rule, e.g. 
\begin{eqnarray}
D(\Phi^{ABC\dots}\wedge\Psi^{DEF\dots})=D\Phi^{ABC\dots}\wedge\Psi^{DEF\dots}+(-1)^p \Phi^{ABC\dots}\wedge D\Psi^{DEF\dots}
\end{eqnarray}
where $\Phi^{ABC\dots}$ is some Lie-algebra-valued p-form, and the gauge covariant exterior derivative reduces to the ordinary exterior derivative for a form with no free gauge indices, e.g.
\begin{eqnarray}
D\Phi^{A}_{\ph AA}=d\Phi^{A}_{\ph AA}
\end{eqnarray}
we can make use of the same tricks as above to vary a Lagrangian four-form which per definition contains no free gauge indices. See Appendix \ref{MMaction} for a concrete example.
\subsection{Topological terms}\label{topologicalterms}
When writing down actions is it important to quickly be able to recognize topological terms since the do not alter the equations of motion. These all have the form $d\Omega$ where $\Omega$ is some three-form. Let $A^{AB}$ and $\omega^{IJ}$ be two connections with $F^{AB}$ and $R^{IJ}$ the corresponding curvature two forms. Two examples of topological terms (i.e. exterior derivatives of three-forms) are then
\begin{align}
F^{AB}\wedge F_{AB}&=d\left(A^{AB}\wedge F_{AB}+\frac{1}{3}A^{AC}\wedge A_{A}^{\phantom{A}D}\wedge A_{CD}\right) \label{top1}\\
\epsilon_{IJKL} R^{IJ}\wedge R^{KL}&=d\left(\epsilon_{IJKL}\omega^{IJ}\wedge( R^{KL}-\frac{1}{3}\omega^{K}_{\ph{K}M}\wedge \omega^{ML})\right) \label{top2}.
\end{align}
Another topological term that includes the contact vector is known as the Nieh-Yan term. We can derive it from the three-form
\begin{eqnarray}
T^A\wedge DV_A\equiv F^{AB}\wedge DV_AV_B.
\end{eqnarray}
by taking its exterior derivative (which is amounts to taking the divergence of its dual)
\begin{eqnarray}\label{nyny}
d(F^{AB}\wedge DV_AV_B)&=&D(F^{AB}\wedge DV_AV_B)=F^{AB}\wedge F_{AC}V^CV_B-F^{AB}\wedge DV_A\wedge DV_B\nonumber\\
&=&T^A\wedge T_A-F^{AB}\wedge DV_A\wedge DV_B\nonumber
\end{eqnarray}
where we have used the identities $DF^{AB}\equiv0$ and $D^2V^A=F^A_{\ph AB}V^B$. Adding the Nieh-Yan term to the Palatini action will not change the equations of motion since it is the exterior derivative of a three-form (or equivalently the divergence of its dual vector density). However, the terms $F^{AB}\wedge DV_A\wedge DV_B$ and $T^A\wedge T_A$ are not topological when taken separately since they are not the exterior derivative of some three-form. However, since their difference is the Nieh-Yan topological term we obtain the same equations of motion if we add either the first or  the second one.

The second term is called the Holst term and as we have just stressed not topological and will therefore yield different equations of motion than the MacDowell-Mansouri action. However, since the Holst term differs from the term $T^A\wedge T_A$ only by the topological Nieh-Yan term, we can add $T^A\wedge T_A$ instead. We can now hope that for vanishing spin-density (which induces torsion) we reproduce General Relativity. Indeed, this is the case as can be verified from the equations of motion. 
\subsection{Example: MacDowell-Mansouri action}\label{MMaction}
As a concrete example of the calculus of variations for forms we consider the MacDowell-Mansouri action with all the essential steps of calculation included. The Bianchi identity $DF^{AB}\equiv0$ simplifies the calculations enormously. As explained in section \ref{nondynapproach}, the normalized and spacelike contact vector $V^A$ is not a dynamical field in the MacDowell-Mansouri action and no variation with respect to it is required. Thus we only consider the variation with respect to $A^{AB}$. Here is the variation of the MacDowell-Mansouri action in pedagogical detail:
\begin{align}
\delta_A {\cal S}_P&=\int_V \delta_A(\epsilon_{ABCDE}V^E F^{AB}\wedge F^{CD})=\int_V \delta_A(\epsilon_{ABCDE}V^E DA^{AB}\wedge DA^{CD})\nonumber\\
&=\int_V \epsilon_{ABCDE}V^E (D\delta A^{AB}\wedge DA^{CD}+DA^{AB}\wedge D\delta A^{CD})=2\int_V \epsilon_{ABCDE}V^E D\delta A^{AB}\wedge F^{CD}\nonumber\\
&=2\int_V D(\epsilon_{ABCDE}V^E \delta A^{AB}\wedge F^{CD})+\delta A^{AB}\wedge D(\epsilon_{ABCDE}V^E F^{CD})\nonumber\\
&=2\int_V d(\epsilon_{ABCDE}V^E \delta A^{AB}\wedge F^{CD})+\delta A^{AB}\wedge \epsilon_{ABCDE} (DV^E\wedge F^{CD}+V^E \underbrace{DF^{CD}}_{\equiv0})\nonumber\\
&=2\underbrace{\int_{\partial V} \epsilon_{ABCDE}V^E \delta A^{AB}\wedge F^{CD}}_{=0}+\int_V\delta A^{AB}\wedge \epsilon_{ABCDE} DV^E\wedge F^{CD}\nonumber\\
&=2\int_V\delta A^{AB}\wedge (\epsilon_{ABCDE} DV^E\wedge F^{CD})\nonumber\\
\end{align}
from which the equations of motions, which naturally appear as a set of three-forms, are readily identified as
\begin{eqnarray}\label{MMwaywiserfieldeq}
\epsilon_{ABCDE} DV^E\wedge F^{CD}=0.
\end{eqnarray}
This equation is nothing but the Palatini-Einstein field equations with positive cosmological constant and zero-torsion condition but written in a compact way; something which the contact vector $V^A$ allows for. 
\section{Einstein equations in standard form}\label{standardform}
Although the field equations (\ref{MMwaywiserfieldeq}) written in waywiser variables are simple and elegant, it is instructive to rewrite it so that they take on the standard more complicated form which we recognize from text books. If we impose the gauge choice $V^A=\ell \delta^A_4$ we have
\begin{eqnarray}
e^I=DV^I\qquad F^{I4}=\mp\frac{1}{\ell} T^I\qquad F^{IJ}=R^{IJ}\pm\frac{1}{\ell^2}e^I\wedge e^J 
\end{eqnarray}
with the sign as prescribed in section \ref{antiwheel}.
\subsection{Einstein field equations}
If we set $A=4$ $B=I$ in equation \eqref{MMwaywiserfieldeq} we get
\begin{eqnarray}
0&=&\epsilon_{4ICDE} DV^E\wedge F^{CD}=\epsilon_{IJKL} e^L\wedge (R^{JK}\pm\frac{1}{\ell^2}e^J\wedge e^K)\nonumber\\
&=&\epsilon_{IJKL} (e_\mu^L dx^\mu)\wedge\left((\frac{1}{2}R_{\nu\rho}^{\ph{\nu\rho}JK}dx^\nu\wedge dx^\rho)\pm\frac{1}{\ell^2}(e_\nu^Jdx^\nu)\wedge(e_\rho^K dx^\rho)\right)\nonumber\\
&=&\epsilon_{IJKL} e_\mu^L (\frac{1}{2}R_{\nu\rho}^{\ph{\nu\rho}JK}\pm\frac{1}{\ell^2}e_\nu^Je_\rho^K)dx^\mu\wedge dx^\nu\wedge dx^\rho
\end{eqnarray}
From this three-form we can construct the dual vector density $\epsilon_{IJKL} e_\mu^L (\frac{1}{2}R_{\nu\rho}^{\ph{\nu\rho}JK}\pm\frac{1}{\ell^2}e_\nu^J e_\rho^K)\varepsilon^{\mu\nu\rho\sigma}$ which after some rewriting is identified as the standard Einstein field equations:
\begin{align}
0&=\varepsilon^{\mu\nu\rho\sigma}\epsilon_{IJKL} e_\nu^L(\frac{1}{2}R_{\rho\sigma}^{\ph{\rho\sigma}JK}\pm\frac{1}{\ell^2}e_\rho^J e_\sigma^K)=\varepsilon^{\mu\nu\rho\sigma}ee^\alpha_Ie_J^\beta e_K^\gamma e_L^\delta\epsilon_{\alpha\beta\gamma\delta} e_\nu^L(\frac{1}{2}R_{\rho\sigma}^{\ph{\rho\sigma}JK}\pm\frac{1}{\ell^2}e_\rho^J e_\sigma^K)\nonumber\\
&=e\varepsilon^{\mu\nu\rho\sigma}e^\alpha_Ie_J^\beta e_K^\gamma \epsilon_{\alpha\beta\gamma\delta}\delta^\delta_\nu(\frac{1}{2}R_{\rho\sigma}^{\ph{\rho\sigma}JK}\pm\frac{1}{\ell^2}e_\rho^J e_\sigma^K)=\frac{e}{2}\varepsilon^{\mu\rho\sigma\nu}\epsilon_{\alpha\beta\gamma\nu}e^\alpha_I R_{\rho\sigma}^{\ph{\rho\sigma}\beta\gamma}\pm \frac{e}{\ell^2}\varepsilon^{\mu\rho\sigma\nu}\epsilon_{\alpha\rho\sigma\nu}e^\alpha_I\nonumber\\
&=\frac{e}{2}(\delta^\mu_\alpha\delta^\rho_\beta\delta^\sigma_\gamma+\delta^\mu_\gamma\delta^\rho_\alpha\delta^\sigma_\beta+
\delta^\mu_\beta\delta^\rho_\gamma\delta^\sigma_\alpha-\delta^\mu_\alpha\delta^\rho_\gamma\delta^\sigma_\beta-
\delta^\mu_\gamma\delta^\rho_\beta\delta^\sigma_\alpha-\delta^\mu_\beta\delta^\rho_\alpha\delta^\sigma_\gamma)e^\alpha_I R_{\rho\sigma}^{\ph{\rho\sigma}\beta\gamma}\pm \frac{e}{\ell^2}3!\delta^\mu_\alpha e^\alpha_I\nonumber\\
&=ee^\alpha_I(\delta_\alpha^\mu R_{\beta\gamma}^{\ph{\beta\gamma}\beta\gamma}+R_{\alpha\beta}^{\ph{\alpha\beta}\beta\mu}+R_{\beta\alpha}^{\ph{\beta\alpha}\mu\beta})\pm \frac{6e}{\ell^2}e^\mu_I=-2e(R_I^{\ph I\mu}-\frac{1}{2}e^\mu_I R\mp \frac{3}{\ell^2}e^\mu_I)\nonumber
\end{align}
where $R\equiv R_{\mu\nu}^{\ph{\mu\nu}IJ}e^\mu_Ie^\nu_J$ and $R_\mu^{\ph \mu I}\equiv R_{\mu\nu}^{\ph{\mu\nu}IJ}e^\nu_J$. We can finally rewrite the equation as 
\begin{eqnarray}
R_\mu^{\ph \mu\nu}-\frac{1}{2}\delta_\mu^\nu R\mp\frac{3}{\ell^2}\delta_\mu^\nu=0
\end{eqnarray}
which is nothing but Einstein field equations with a positive cosmological constant $\Lambda=\mp\frac{3}{\ell^2}$.
\subsection{Vanishing torsion}
To demonstrate that the torsion tensor vanishes we set $A=I$ and $B=J$ in equation \eqref{MMwaywiserfieldeq} which yields:
\begin{eqnarray}
0=\epsilon_{IJ4KL} e^L\wedge F^{4K}=\frac{1}{\ell}\epsilon_{IJKL} e^L\wedge T^K=\frac{1}{2\ell}\epsilon_{IJKL} e_\mu^L T_{\nu\rho}^Kdx^\mu\wedge dx^\nu\wedge dx^\rho
\end{eqnarray}
To see what this means in tensor language we rewrite the three-form as a dual vector density $\epsilon_{IJKL} e_\mu^L T_{\nu\rho}^K\varepsilon^{\mu\nu\rho\sigma}$. The steps are similar to the rewriting of the Einstein field equations and we do not display calculation in detail. The result is:
\begin{eqnarray}\label{torsiontensoreq}
\epsilon_{IJKL} e_\mu^L T_{\nu\rho}^K\varepsilon^{\mu\nu\rho\sigma}=-e(T^\sigma_{IJ}+e^\sigma_I T_{J\mu}^\mu-e^\sigma_J T_{I\mu}^\mu)=0.
\end{eqnarray}
Contracting this equation with $e_\sigma^J$ yields $T_{I\mu}^\mu=0$ which when inserted back into \eqref{torsiontensoreq} yields $T^\sigma_{IJ}=0$. Thus, the equations of motion imposes zero torsion which shows that the MacDowell-Mansouri action is equivalent to the Einstein-Hilbert action.
\section{Bibliography}
\bibliographystyle{hunsrt}
\bibliography{references}
\end{document}